**The UPLNC Compiler: Design and Implementation**

Evgueniy Vitchev[1]

Department of Physics and Astronomy, Rutgers University
Piscataway, NJ 08855-0849, USA

**Abstract**

The implementation of the compiler of the UPLNC language is presented with a full source code listing.

# 1 Introduction

## 1.1 Why develop another language?

There were several motivations to develop the UPLNC language. One of them was to make a programming language which suits better one's personal tastes, and allows experimenting with ideas. Also, developing of the compiler was an amusing process for the author, who hopes that this work will be instructive to the reader. From a certain point, one could view such a work as a work of art, like a poet would write a beautiful poem. Another reason behind this language would be the beginning of a much larger project- building an alternative do-it-yourself (DIY) platform, including programming language, developer tools, operating system kernel written in that language, full set of system and application utilities. If some day such a project becomes implemented, it is the author's hope that it will remain free, as free speech, without being threatened by claims of exclusive rights to use, modify and distribute. That's why the source code in the present work is released under the GNU General Public License.

This language is already quite developed- the compiler does support unlimited level of indirection, multiple dimensional arrays, structures, methods of structures, and most of the control statements inherent of C. On the other hand, there are some more or less essential things which would be desirable to be implemented, e.g. floating point support, initializers and optimization phase before the code generation.

An open question is whether this language will evolve as a general-purpose one like C and C++, or will have some more sophisticated features as combinatoric control statements(CCS). An lyrical diversion which gives some notion of CCS is a calculation of the form

$$R = \sum_{1 \le i_1 \le i_2 \le \ldots \le i_k \le N} f(i_1, i_2, \ldots, i_k). \tag{1}$$

When $k$ (the number of indices) is not known, the simplest way is to write some recursive or iterative algorithm which cycles the $k$ indices in the appropriate way. This can be done also with a standard library function/template. A non-trivial way to deal with this situation is post-compile code generation: the program can generate machine code corresponding to something like

---

[1]e-mail: vitchev@physics.rutgers.edu

1

```
for(i[k-1]=1;i[k-1]<=N;++i[k-1])
 for(i[k-2]=1;i[k-2]<=i[k-1];++i[k-2])
  for(i[k-3]=1;i[k-3]<=i[k-2];++i[k-3])
   ...
      R+=f(i);
```

and run that machine code, or it can output the source code to a file, invoke an external compiler which would produce a shared object, which then would be loaded and run.

## 1.2   Development

In part, this work was inspired by Small C compiler [1] and its variant RatC [3]. When the author encountered RatC around one decade ago, he made a rough implementation(or translation) of it in Turbo Pascal. That project was ultimately abandoned, mainly because it was a compiler written in a quite high-level language, translating(by design) a rather primitive language, so a bootstrap procedure was not possible. Its source code is probably lost.

Much later, when I saw several other implementations of Small C, decided that it would be better, instead of making subsets and dialects, which will probably never be as good as the complete language, to make one's own language, referred to as UPLNC, which will allow more choices of its design. Still, I decided to use Small C as the first language to implement the UPLNC compiler– to that end I ported a version of Small C to `i386-linux` [4]. Those preliminary versions of the compiler were pretty much constrained by the paradigms of the Small C compiler– no use of structures at all, so that tables were implemented by means of several arrays, corresponding to the different fields of the missing structures. Even in the present version one can find similarities of the syntax parser of UPLNC to the one of Small C, as well as artefacts of code borrowed from the Small C compiler. Nevertheless, the author wanted to escape the constraints of Small C, that's why, when the compiler was already able to produce correct code, the author wrote a utility for translating Small C to UPLNC (one can see its traces by some truncated identifiers like `expressi`). The result was a compiler for UPLNC, written in the same language, so it could compile itself.

A next step was to replace the implementation of the tables as arrays of scalar types with arrays or lists of structures. The processing of expressions was completely rewritten. The old Small C-style approach of code generation being done while parsing, was replaced by a three-stage process: parsing and building a tree, generating from that three an intermediate representation of the program code, producing of assembly code. It is possible in the future to include an optimization stage after the second stage of this process.

Later, methods of structures were introduced in the language, much similar to the methods in C++: the method has a default parameter `this` which points to the instance of the structure, and within the method the members of the structure are visible.

It is possible to declare variables in different styles:

```
var a,b,c:[3]*char;
var [3]*char:d,e,f;
var extern a1,a2:int;
var extern int:a3;
var int extern:a4;
var a5:extern int;
var a6:int extern;
var extern a7:extern int extern;
var extern **int extern: extern a8;
```

Currently, the only supported target of the compiler is Linux on i386-compatible processors. The code generated is link-compatible with C, that's why it is possible to link programs with the GNU C library `glibc`, which is used by the compiler itself. It is plausible that with small modifications the compiler could support other operating systems running on `i386`-compatible microprocessors. In order to support other architectures, it is possible to rewrite the code generating part of the compiler. This was the main reason behind separating the code generation in a different stage and even in the different module `codegen.e`.

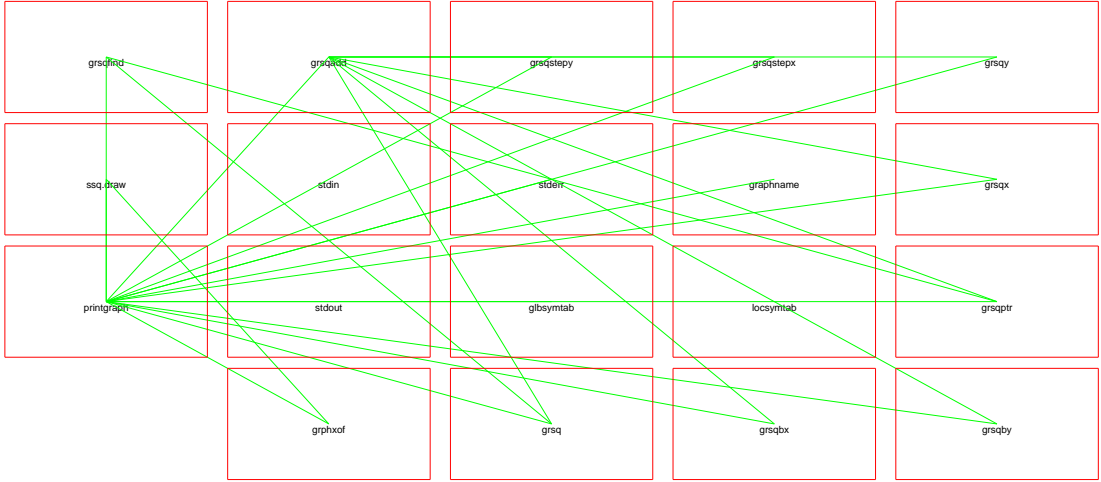

Figure 1: Scheme of `grph.e`

A feature introduced in the compiler was the possibility of redefining certain syntax elements. An illustrative, albeit not quite readable example is this program for calculating of primes:

```
%%takeoff lcomp "<font>" ;
%%takeoff rcomp "</font>" ;
%%takeoff tfunc "<p>" ;
%%takeoff larg "<arg>" ;
%%takeoff rarg "$</arg>" ;

#define MAXNUM 1000
#define TABSZ  1001

var tab:[TABSZ]int;/*factorizes?*/

<p> doprimes<arg>$</arg>
<font>
  var int:i,k;
  printf<arg>"Calculating the primes>=3...\n"$</arg>;
  for<arg>i=2;i<TABSZ;i++$</arg>
    tab[i]=0;
  for<arg>i=3;i<TABSZ;i=i+2$</arg>
    <font>
      if<arg>tab[i]$</arg>
        continue;
      printf<arg>"%d\n",i$</arg>;
```

```
        for<arg>k=i;k<TABSZ;k=k+i$</arg>
            <font>
                tab[k]=1;
            </font>
        </font>
</font>
<p> main<arg>$</arg>
<font>
    doprimes<arg>$</arg>;
</font>
```

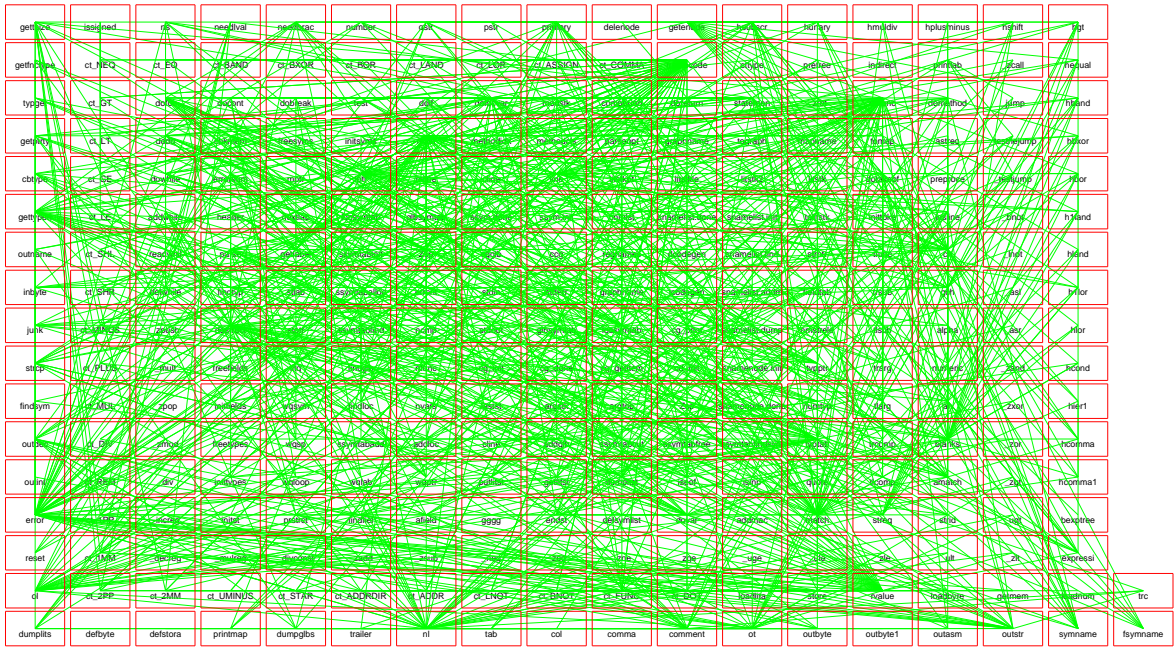

Figure 2: Scheme of `langc.e`

Another feature which was implemented lately was the code graphing, namely producing a `gnuplot` [6] input file containing a graphical scheme of the referrals between global symbol names in a particular module. For example, the scheme produced for some of the modules in the compiler are shown in Fig.1 and Fig.2.

The compiler driver was written in Perl [8]. It runs the preprocessor `lpp1` and redirects its output as input of the compiler `langc`. Linking is done by using the driver of `gcc` [7].

## 2   Source

In this section a full source code listing of the compiler is presented. An archive of previous versions is in [5].

## 2.1  langc.e

```
/*                    -*- C -*-                                    */
/*            The language compiler by E.V., (C) 2003             */
/*
  This program is free software; you can redistribute it and/or modify
  it under the terms of the GNU General Public License as published by
  the Free Software Foundation; either version 2 of the License, or
  (at your option) any later version.

  This program is distributed in the hope that it will be useful,
  but WITHOUT ANY WARRANTY; without even the implied warranty of
  MERCHANTABILITY or FITNESS FOR A PARTICULAR PURPOSE.  See the
  GNU General Public License for more details.

  You should have received a copy of the GNU General Public License
  along with this program; if not, write to the Free Software
  Foundation, Inc., 59 Temple Place, Suite 330, Boston, MA  02111-1307  USA
*/
#include "tlangc.he"
#include "codegen.he"

var errcnt:int;
var ncmp:int;
var nfunc:int;
var nvars:int;
var lastst:int;
var argstk:int;
var argtop:int;
var Zsp:int;
method snamenode.done()
{
  if(next)
  {
    next->done();
    free(next);
  }
}
method snamenode.init(){next=0;name[0]=0;}
method snamelist.dump()
{
  var *snamenode:p;
  for(p=lst;p;p=p->next)
  {
    fprintf(stderr,"%s\n",p->name);
  }
}
method snamelist.addm(*char:s)/*add(merge)*/
{
```

```
    if(!this)return;
    var *snamenode:p;
    if(p=find(s))
    return;
    chkmem(*front=calloc(1,sizeof(snamenode)));
    (*front)->init();
    strncpy((*front)->name,s,NAMESIZE);
    front=&(*front)->next;
}
method snamelist.find(*char:s)
{
    var *snamenode:p;
    for(p=lst;p;p=p->next)
    if(strid(p->name,s))
        return p;
    return 0;
}
method snamelist.init(){lst=0;front=&lst;}
method snamelist.done()
{
    front=&lst;
    if(lst)
    lst->done();
}
var *snamelist :cnmlst;
method ssym.init()
{
    chkmem(nmlst=calloc(1,sizeof(snamelist)));
    nmlst->init();
}
method ssym.done()
{
    if(nmlst)
    {
        nmlst->done();
        free(nmlst);
    }
}
var glbsymtab,locsymtab:ssymtab;
func ssymtabinit(p:*ssymtab)
{
    p->lst=0;
    p->front=&p->lst;
}
func ssymtaballoc(p:*ssymtab)
{
    var *ssym:res;
    (*p->front)=chkmem(calloc(1,sizeof(ssymlist)));
    (*p->front)->next=0;
```

```
  res=&(*p->front)->sym;
  p->front=&(*p->front)->next;
  return res;
}
func ssymtabfind(p:*ssymtab,name:*char)
{
  var *ssymlist:q;
  for(q=p->lst;q;q=q->next)
  {
    if(strid(name,q->sym.name))
    return &q->sym;
  }
  return 0;
}
func findglb(name:*char)
{
  return ssymtabfind(&glbsymtab,name);
}
func findloc(name:*char)
{
  return ssymtabfind(&locsymtab,name);
}
func ssymtabadd(p:*ssymtab,sname:*char,sort:int,dfd:int,offset:int,type:int)
{
  var *ssym:sym;
  sym=ssymtaballoc(p);
  sym->init();
  sym->sort=sort;
  sym->dfd=dfd;
  sym->offset=offset;
  sym->type=type;
  strcp(sym->name,sname);
  return sym;
}
func addloc(sname:*char,sort:int,dfd:int,offset:int,type:int)
{
  return ssymtabadd(&locsymtab,sname,sort,dfd,offset,type);
}
var extern cline:int;
func addglb(sname:*char,sort:int,dfd:int,offset:int,type:int)
{
  var *ssym:res;
  res=ssymtabadd(&glbsymtab,sname,sort,dfd,offset,type);
  res->line=cline;
  return res;
}
func ssymtabcut(p:*ssymtab,w:**ssymlist)
{
  var *ssymlist:i,n;
```

```
  for(i=*w;i;i=n)
  {
    n=i->next;
    free(i);
  }
  *w=0;
  p->front=w;
}
func ssymtabfree(p:*ssymtab)
{
  var *ssymlist:q;
  var *ssymlist:t;
  for(q=p->lst;q;q=t)
  {
    t=q->next;
    q->sym.done();
    free(q);
  }
  p->lst=0;
  p->front=&p->lst;
}
func ssymtabfindsym(symtab:*ssymtab,sname:*char)
{
  var *ssymlist:p;
  for(p=symtab->lst;p;p=p->next)
  if(strid(sname,p->sym.name))
    return p;
  return 0;
}
var typtab:*styp;
var numtyp:int;
/*var typnames:[TYPNMTBS]char;*/
/*var typsort:[NUMTYP]int;*/
/*var typtype:[NUMTYP]int;*/
/*var typdim:[NUMTYP]int;*/
/*var typsize:[NUMTYP]int;*/
var typptr:int;
var nmstrele:int;
var fieldtab:*sfield;
/*var strnames:[STRNMTBS]char;*/
/*var strtyp:[NMSTRELE]int;*/
/*var stroffse:[NMSTRELE]int;*/
/*var strnext:[NMSTRELE]int;*/
var strptr:int;
var tolitstk:int;
var litstk:[litstksz]char;
var litstk2:[litstksz]char;
var litstkle:[litstknu]int;
var litstkpt:[litstknu]int;
```

```
var line:[linesize]char;
var cline:int;
var mline:[linesize]char;
var lptr:int;
var mptr:int;
var nextlab:int;
func getlabel()
{
  return ++nextlab;
}
var stlab:int;
var stptr:int;
var litq:[STSIZE]char;
var wqsym:[WQNUM]int;
var wqsp:[WQNUM]int;
var wqloop:[WQNUM]int;
var wqlab:[WQNUM]int;
var wqptr:int;

func putlitst(c:char)
{
  if(litstkpt[tolitstk]+litstkle[tolitstk]>=litstksz-1){
  error("too large code from function arguments");
  return 0;
  }
  litstk[litstkpt[tolitstk]+litstkle[tolitstk]]=c;
  litstkle[tolitstk]=litstkle[tolitstk]+1;
  return c;
}
func getlitst()
{
  if(tolitstk>=litstknu-1){
  error("too many function arguments");
  return 0;
  }
  ++tolitstk;
  litstkle[tolitstk]=0;
  litstkpt[tolitstk]=litstkpt[tolitstk-1]+litstkle[tolitstk-1];
  return tolitstk;
}
func dumpltst(tl:int)
{
  var int:i;
  var int:p;
  var *char:q;
  var *char:pp;
  q=litstk2;
```

```
  while(tolitstk>=tl){
  i=litstkle[tolitstk];
  p=litstkpt[tolitstk];
  while(i--){
     *q++=litstk[p++];
  }
  tolitstk--;
  }
  pp=q;
  q=litstk2;
  while(q<pp){
  outbyte(*q);
  ++q;
  }
}
var iseof:int;
var isinp:int;
var quote:[2]char;
var tlcomp:[NAMESIZE]char;
var trcomp:[NAMESIZE]char;
var tlarg:[NAMESIZE]char;
var trarg:[NAMESIZE]char;
var tlsub:[NAMESIZE]char;
var trsub:[NAMESIZE]char;
var tfunc:[NAMESIZE]char;
func inittoke()
{
  strcp(tlcomp,"{");
  strcp(trcomp,"}");
  strcp(tlarg,"(");
  strcp(trarg,")");
  strcp(tlsub,"[");
  strcp(trsub,"]");
  strcp(tfunc,"func");
}
func dotakeof()
{
  var *char:token;
  var int:k,c;
  if(amatch("lcomp",5))token=tlcomp;
  else if(amatch("rcomp",5))token=trcomp;
  else if(amatch("larg",4))token=tlarg;
  else if(amatch("rarg",4))token=trarg;
  else if(amatch("lsub",4))token=tlsub;
  else if(amatch("rsub",4))token=trsub;
  else if(amatch("tfunc",5))token=tfunc;
  else
  {
    error("token name expected");
```

```
    return;
  }
  blanks();
  if(ch()!='"')
  {error("qstr expected");return;}
  gch();
  k=0;
  while(ch()!='"')
  {
    if(!ch())break;
    c=gch();
    if(c==92)
    {
      c=gch();
      if(!c)break;
      if(c=='n')c=10;
      else if(c=='t')c=9;
      else if(c=='b')c=8;
      else if(c=='f')c=12;
    }
    if(k<NAMEMAX)
    token[k++]=c;
  }
  if(ch()=='"')gch();
  token[k]=0;
  comment();outstr("new token:");outstr(token);nl();
  ns();
}
var int:tomap;
var *char:mapname;
var int:tograph;
var *char:graphname;
func parseopt(argc:int,argv:**char)
{
  tomap=0;
  mapname="tlmap.map";
  tograph=0;
  graphname="tlgraph.dat";
  /*fprintf(stderr,"parseopt()\n");*/
  var int:i;
  for(i=1;i<argc;i++)
  {
    if(*argv[i]=='-')
    {
      fprintf(stderr,"option:\n");
      var k:int;
      for(k=1;argv[i][k];k++)
      if(argv[i][k]=='m')
        {
```

```
          tomap=1;
          if(argv[i][k+1]=='-')
            {tomap=0;k++;}
          fprintf(stderr," tomap=%d\n",tomap);
          }
      else if(argv[i][k]=='g')
        {
        tograph=1;
        if(argv[i][k+1]=='-')
          {tograph=0;k++;}
        fprintf(stderr," tograph=%d\n",tograph);
        }
      else if(argv[i][k]=='M')
        {
        if(argv[i][k+1])
          {
          mapname=argv[i]+k+1;
          while(argv[i][k])++k;
          }
        else if(i>=argc-1)
          {
          fprintf(stderr,"needs more arguments\n");
          }
        else
          {
          mapname=argv[++i];
          }
        fprintf(stderr," mapname=%s\n",mapname);
        }
      else if(argv[i][k]=='G')
        {
        if(argv[i][k+1])
          {
          graphname=argv[i]+k+1;
          while(argv[i][k])++k;
          }
        else if(i>=argc-1)
          {
          fprintf(stderr,"needs more arguments\n");
          }
        else
          {
          mapname=argv[++i];
          }
        fprintf(stderr," graphname=%s\n",graphname);
        }
  }
 }
}
```

```
var methodcls:int;/*whether we are in a method*/
var methodidx:int;
/*var methodths:*ssym;*/
/*var methodstr:int;*///*it's a type!*/
func main(argc:int,argv:**char)
{
  tolitstk=litstkpt[0]=litstkle[0]=ncmp=lastst=nextlab=
  lptr=iseof=strptr=stptr=errcnt=argstk=wqptr=line[0]=quote[1]=0;
  nfunc=0;nvars=0;cline=0;
  typptr=F_TYPE;
  strptr++;
  isinp=1;
  cnmlst=0;
  /*glbptr++;*/
  *quote='"';
  methodcls=0;methodidx=0;/*methodstr=0;*/
  initdyn();
  var *char:pp,pp2;pp=autodynstr("Hello ");pp2=autodynstr("world\n");
  stlab=getlabel();
  parseopt(argc,argv);
  initsyms();
  inittoke();
  inittypes();
  initfields();
  initst();
  icodegen();
  header();
  parse();
  dumplits();
  dumpglbs();
  if(tomap)
  printmap();
  if(tograph)
  printgraph();
  trailer();
  dcodegen();
  errorsum();
  freefields();
  freetypes();
  freesyms();
  /*fprintf(stderr,"(%d)(%d)%s%s",pp,pp2,pp,pp2);*/
  donedyn();
  return 0;
}
func initsyms()
{
  ssymtabinit(&glbsymtab);
  ssymtabinit(&locsymtab);
}
```

```
func freesyms()
{
  ssymtabfree(&locsymtab);
  ssymtabfree(&glbsymtab);
}
func chkmem(p:*int)
{
  /*fprintf(stderr,"chkmem(%d)\n",p);*/
  if(!p)
  {
    error("out of memory");
    exit(2);
  }
  else
  return p;
}
func errorsum()
{
  if(ncmp)error("missing '}'");
  comment();
  outstr(" ");
  outdec(errcnt);
  outstr(" error(s)");
  nl();
  comment();
  outstr(" ");
  outdec(nfunc);
  outstr(" function(s), ");
  outdec(nvars);outstr(" global variables");
  nl();
}
func header()
{
  comment();
  outstr("The Test Compiler");
  nl();
}
func parse()
{
  while(!iseof)
  {
    if(match("#define"))addmac();
    else if(amatch("%%takeoff",9))
    {
      comment();
      outstr(">>>>");
      nl();
      dotakeof();
    }
```

```
    else if(amatch("var",3))dovar();
    else if(amatch("struct",6))prstrct(dostruct());
    else if(amatch(tfunc/*"func"*/,4))dofunc();
    else if(amatch("method",6))domethod();
    else
    {
      error("wrong declaration- top level");
      reset();
    }
    blanks();
  }
}
func findtyp(sname:*char)
{
  var int:res;
  var *char:p;
  res=1;
  while(res<typptr){
  p=typtab[res].name;
  if(astreq(sname,p,NAMEMAX))return res;
  res++;
  }
  return 0;
}
func dostruct()
{
  var [NAMESIZE]char:sname;
  var [NAMESIZE]char:fname;
  var int:t;
  var int:sz;
  /*var *int:top;*/
  var int:ttop;
  var int:offs;
  var int:res;
  if(typptr>=numtyp){
  /*fprintf(stderr,"reallocating types\n");*/
  numtyp=numtyp+NUMTYP;
  chkmem(typtab=realloc(typtab,sizeof(styp)*numtyp));
  }
  res=typptr++;
  if(!symname(sname))error("structure name exected");
  if(findtyp(sname))error("multidef type");
  comment();
  outstr("name of struct:");
  outstr(sname);
  nl();
  comment();
  outstr("1typptr=");
  outdec(typptr);
```

```
nl();
strcp(typtab[res].name,sname);
typtab[res].sort=V_STR;
needbrac(tlcomp/*"{"*/);
ttop=0;/*top=&typtab[res].type*//*typtype+res*/;
/* *top=0;*/typtab[res].type=0;
offs=0;
while(1){
blanks();
if(iseof)break;
/*if(ch()=='}')break;*/
if(streq(line+lptr,trcomp))break;
if(amatch(tfunc/*func*/,4))
    {
    if(cbtype())
      t=gettypen();
    else t=T_INT;
    t=getfnctype(t);
    if(!symname(fname))
      error("method name expected");
    ns();
    if(!ttop)
      {
      comment();outstr("BEGINNING\n");
      typtab[res].type=afield(t,fname,offs);
      ttop=typtab[res].type;
      }
    else
      {
      comment();outstr("assigning");
      fieldtab[ttop].next=afield(t,fname,offs);
      nl();
      ttop=fieldtab[ttop].next;
      }
    /**top=afield(t,fname,offs);*/
    /*top=&fieldtab[*top].next;*/
    /*sz=gettsize(t);
    if(sz&3)sz=sz+4-(sz&3);
    offs=offs+sz;*/
    continue;
    }
var int:istyp;
istyp=0;
if(cbtype())
  {
  t=gettypen();
  istyp=1;
  if(match(":"));
  }
```

```
var *ssymlist:lst,lsptr;
var **ssymlist:lpom;
lst=lsptr=0;
lpom=&lst;
while(symname(fname))
   {
   chkmem((*lpom)=calloc(1,sizeof(ssymlist)));
   strcp((*lpom)->sym.name,fname);
   lpom=&(*lpom)->next;
   (*lpom)=0;
   blanks();
   if(istyp)
      {if(streq(line+lptr,";"))break;}
   else
      {if(streq(line+lptr,":"))break;}
   if(endst())break;
   if(!match(","))
      {
      error("',' or ':'|';' expected");
      junk();
      break;
      }
   }
if(!istyp)
   {
   if(!match(":"))
      error("':' expected for Pascal-style declaration");
   t=gettypen();
   istyp=1;
   }
ns();
/*if(!symname(fname)){
   error("field name expected");
   reset();
   }*/
for(lsptr=lst;lsptr;lsptr=lsptr->next)
   {
   comment();outstr("adding ");outstr(lsptr->sym.name);
   outstr(":");outdec(t);nl();
   if(!ttop)
      {
      typtab[res].type=afield(t,lsptr->sym.name,offs);
      ttop=typtab[res].type;
      }
   else
      {
      var int:alpha;
      comment();outstr("ttop=");outdec(ttop);nl();
      alpha=afield(t,lsptr->sym.name,offs);
```

```
        fieldtab[ttop].next=gggg();
        comment();outstr("fieldtab[ttop].next=");
        outdec(fieldtab[ttop]);nl();
        fieldtab[ttop].next=alpha;
        /*fieldtab[ttop].next=afield(t,lsptr->sym.name,offs);*/
        comment();outstr("fieldtab[ttop].next=");
        outdec(fieldtab[ttop].next);
        ttop=fieldtab[ttop].next;
        }
     comment();outstr("ttop=");outdec(ttop);nl();
     prstrct(res);
     /**top=afield(t,lsptr->sym.name,offs);
     comment();outstr("*top=");outdec(*top);nl();
     top=&fieldtab[*top].next;*/
     sz=gettsize(t);
     if(sz&3)sz=sz+4-(sz&3);
     offs=offs+sz;
     }
  delsymlist(lst);
  }
  comment();
  outstr("2typptr=");
  outdec(typptr);
  nl();
  typtab[res].size=offs;
  needbrac(trcomp/*"}"*/);
  ns();
  comment();
  outstr("typptr=");
  outdec(typptr);
  nl();
  return res;
}
func freefields()
{
  if(fieldtab)
  free(fieldtab);
}
func initfields()
{
  nmstrele=NMSTRELE;
  chkmem(fieldtab=calloc(nmstrele,sizeof(sfield)));
}
func freetypes()
{
  if(typtab)
  free(typtab);
}
func inittypes()
```

```
{
  numtyp=NUMTYP;
  chkmem(typtab=calloc(numtyp,sizeof(styp)));
  strcp(typtab[T_INT].name,"int");
  typtab[T_INT].sort=V_FND;
  typtab[T_INT].size=WORDSIZE;
  strcp(typtab[T_CHAR].name,"char");
  typtab[T_CHAR].sort=V_FND;
  typtab[T_CHAR].size=BYTESIZE;
  typtab[T_INTP].name[0]=0;
  typtab[T_INTP].sort=V_PTR;
  typtab[T_INTP].type=T_INT;
  typtab[T_INTP].size=WORDSIZE;
  typtab[T_CHARP].name[0]=0;
  typtab[T_CHARP].sort=V_PTR;
  typtab[T_CHARP].type=T_CHAR;
  typtab[T_CHARP].size=WORDSIZE;
}
func initst()
{
}
func prstrct(s:int)
{
  var int:k;
  comment();
  outstr("structn ");
  outdec(s);
  outstr(" sz:");
  outdec(typtab[s].size);
  outstr(" name ");
  outstr(typtab[s].name);
  nl();
  k=typtab[s].type;
  while(k){
  comment();
  outstr("type ");
  outdec(fieldtab[k].type);
  outstr(" ");
  outstr(fieldtab[k].name);
  nl();
  k=fieldtab[k].next;
  }
}
func findfiel(nm:*char,s:int)
{
  var int:k;
  comment();
  outstr("typsort[");
  outdec(s);
```

```
    outstr("]=");
    outdec(typtab[s].sort);
    nl();
    if(typtab[s].sort!=V_STR){
    error("cannot find a field in a non-struct");
    return 0;
    }
    k=typtab[s].type;
    while(k){
    if(strid(nm,fieldtab[k].name))return k;
    k=fieldtab[k].next;
    }
    return 0;
}
func afield(t:int,fname:*char,offs:int)
{
    /*fprintf(stderr,"strptr=%d\n",strptr);*/
    comment();outstr("strptr=");outdec(strptr);nl();
    if(strptr>=nmstrele)
    {
        /*fprintf(stderr,"reallocating field table\n");*/
        comment();outstr("reallocating field table");nl();
        nmstrele=nmstrele+NMSTRELE;
        comment();outstr("nmstrele=");outdec(nmstrele);nl();
        chkmem(fieldtab=realloc(fieldtab,nmstrele*sizeof(sfield)));
    }
    strcp(fieldtab[strptr].name,fname);
    fieldtab[strptr].type=t;
    fieldtab[strptr].offset=offs;
    fieldtab[strptr].next=0;
    comment();outstr("++,strptr=");outdec(strptr);nl();
    return strptr++;
}
func gggg(){return 209;}
func endst()
{
    blanks();
    return ((ch()==';')||(ch()==0));
}
func delsymlist(p:*ssymlist)
{
    if(!p)return;
    delsymlist(p->next);
    free(p);
}
func dovar()
{
    var int:typ,istyp,wdfd;
    var *ssymlist:lst,lptr;
```

```
var **ssymlist:lpom;
var *ssym:idx;
var [NAMESIZE]char:sname;
istyp=0;
lst=lptr=0;
lpom=&lst;
wdfd=1;
while(1)
{
  if(amatch("extern",6))wdfd=0;
  else break;
}
if(cbtype())
{
  istyp=1;
  typ=gettypen();
  while(1)
  {
    if(amatch("extern",6))wdfd=0;
    else break;
  }
  if(!match(":"))
  error("':' expected");
  while(1)
  {
    if(amatch("extern",6))wdfd=0;
    else break;
  }
}
blanks();
if(!an(ch()))error("expected name of variable");
while(symname(sname))
{
  chkmem((*lpom)=calloc(1,sizeof(ssymlist)));
  strcp((*lpom)->sym.name,sname);
  lpom=&(*lpom)->next;
  (*lpom)=0;
  if(istyp)
  {if(match(";"))break;}
  else
  {if(match(":"))break;}
  if(endst())break;
  if(!match(","))
  {
    error("',' or ':'|';' expected");
    junk();
    break;
  }
}
```

```
  if(!istyp)
  {
    while(1)
     {
      if(amatch("extern",6))wdfd=0;
      else
      break;
     }
    typ=gettypen();
    while(1)
     {
      if(amatch("extern",6))wdfd=0;
      else
      break;
     }
    ns();
  }
  for(lptr=lst;lptr;lptr=lptr->next)
  {
    idx=findglb(lptr->sym.name);
    if(idx)
     {
      if(idx->sort!=S_VARG)
       {
        error("multidef global var");
        break;
       }
      else if(idx->dfd)
       {
        error("multidef global var, was defined");
        break;
       }
      else if(typ!=idx->type)
       {
        error("conflicting types!");
        break;
       }
     }
    addglb(lptr->sym.name,S_VARG,wdfd,0,typ);
    if(wdfd)++nvars;
  }
  delsymlist(lst);
}
func addmac()
{
  comment();
  ol(".mac");
}
func match(lit:*char)
```

```
{
  var int:k;
  blanks();
  if(k=streq(line+lptr,lit)){
  lptr=lptr+k;
  return 1;
  }
  return 0;
}
func streq(str1:*char,str2:*char)
{
  var int:k;
  k=0;
  while(str2[k]){
  if(str1[k]!=str2[k])return 0;
  ++k;
  }
  return k;
}
func strid(str1:*char,str2:*char)
{
  var int:k;
  k=0;
  while(str2[k]){
  if(str1[k]!=str2[k])return 0;
  ++k;
  }
  if(str1[k])return 0;
  return 1;
}
func amatch(lit:*char,len:int)
{
  var int:k;
  blanks();
  if(k=astreq(line+lptr,lit,len)){
  lptr=lptr+k;
  return 1;
  }
  return 0;
}
func blanks()
{
  while(1)
  {
    while(!ch())
     {
       insline();
       preproce();
       if(iseof)break;
```

```
        }
        if(ch()==' ')gch();
        else
        if(ch()==9)gch();
        else
          return ;
    }
}
func an(c:char)
{
    /*return ((alpha(c))||(numeric(c)));*/
    return ((c>='a')&&(c<='z'))||((c>='A')&&(c<='Z'))||
    ((c>='0')&&(c<='9'))||(c=='_');
}
func numeric(c:char)
{
    return ((c>='0')&&(c<='9'));
}
func alpha(c:char)
{
    return (((c>='a')&&(c<='z'))||((c>='A')&&(c<='Z'))||(c=='_'));
}
func gch()
{
    if(!ch())return 0;
    return line[lptr++];
}
func ch()
{
    return line[lptr];
}
func insline()
{
    var int:k;
    if(!isinp)iseof=1;
    if(iseof)return ;
    lptr=0;
    line[0]=0;
    while((k=getchar())>0){
    if((k==10)||(lptr>=linemax))break;
    line[lptr++]=k;
    }
    line[lptr]=0;
    if(k<0)isinp=0;
    if(lptr){
    comment();
    outstr(line);
    nl();
    lptr=0;
```

```
    }
    cline++;
}
func preproce()
{
}
func astreq(str1:*char,str2:*char,len:int)
{
    var int:k;
    k=0;
    while(k<len){
    if((str1[k])!=(str2[k]))break;
    if(!str1[k])break;
    if(!str2[k])break;
    ++k;
    }
    if(an(str1[k]))return 0;
    if(an(str2[k]))return 0;
    return k;
}
func domethod()
{
    methodcls=1;
    var [NAMESIZE]char:tname,mname;
    if(!symname(tname))
    {error("symbol name expected");methodcls=0;return;}
    if(!(match(".")||match("::")))
    {error("'.' or '::' expected");methodcls=0;return;}
    if(!symname(mname))
    {error("method name expected");methodcls=0;return;}
    var int:stridx;
    stridx=findtyp(tname);
    if(!stridx)
    {error("no such type");methodcls=0;return;}
    if(typtab[stridx].sort!=V_STR)
    {error("must be a structure to have a method");methodcls=0;return;}
    var int:k;
    if(!(k=findfiel(mname,stridx)))
    {error("no such method");methodcls=0;return;}
    if(typtab[fieldtab[k].type].sort!=V_FNC)
    {error("method must be a function");methodcls=0;return;}
    methodcls=stridx;/*the type in which the method is*/
    methodidx=k;/*the index of the method in fieldtab*/
    dofunc();
    methodcls=0;
    methodidx=0;
}
func dofunc()
{
```

```
var [NAMESIZE]char:n;
var [NAMESIZE]char:argn;
var int:argtype;
var *ssym:gp;
var int:k;
comment();
ol(".fnc");
if(methodcls&&methodidx)
{
  strcp(n,typtab[methodcls].name);
  strcat(n,".");/* DANGEROUS !! */
  strcat(n,fieldtab[methodidx].name);/* DANGEROUS- FixMe */
}
else if(!symname(n))
{
  error("wrong function declaration");
  reset();
  return ;
}
comment();
outstr(">> ");
outstr(n);
nl();
gp=findglb(n);
/*comment();
outstr("gp:");
outdec(gp);
nl();*/
if(gp){
if(gp->sort!=S_FUNC)error("multidef");
}
else
{
  gp=addglb(n,S_FUNC,0,0,T_INT);
}
if(!match(tlarg/*"("*/))error("missing '('");
argstk=0;
Zsp=0;
/*locptr=STARTLOC;*/
ssymtabfree(&locsymtab);
if(methodcls&&methodidx)
{
  addloc("this",S_VARL,1,argstk+8,getptrty(methodcls));
  argstk=argstk+sizeof(*int);
}
while(!match(trarg/*")"*/))
{
  if(cbtype())
  {
```

```
    argtype=gettypen();
    if(match(":"));
    if(!symname(argn))
    {
      error("parameter name expected");
      junk();
    }
    /*if(endst())break;*/
  }
  else
  {
    if(!symname(argn))
    {
      error("parameter name expected");
      junk();
    }
    if(endst())break;
    if(!match(":"))error("':' expected");
    argtype=gettypen();
  }
  k=gettsize(argtype);
  if(k&3)k=k+4-(k&3);
  addloc(argn,S_VARL,1,argstk+8,argtype);
  argstk=argstk+k;
  if(!match(","))if(ch()!=')')
  {
    error("comma or ')' expected");
    break;
  }
  if(endst())break;
}
if(match(";"))return ;
nfunc++;
var *snamelist:savenmlst;
savenmlst=cnmlst;
cnmlst=gp->nmlst;
var *scodegen:savecg;
/*fprintf(stderr,"dofunc:ccg=%d\n",ccg);*/
savecg=ccg;
/*fprintf(stderr,"dofunc:savecg=%d\n",savecg);*/
var scodegen:codeg;
cg_init(&codeg);
ccg=&codeg;
if(gp->dfd)error("this function was already defined");
gp->dfd=1;
ol(".text");
ol(".align 16");
outasm(".globl ");
outname(n);
```

```
   nl();
   ot(".type");
   tab();
   outname(n);
   outasm(",@function");
   nl();
   outname(n);
   col();
   nl();
   /*ol("pushl %ebp");
   ol("movl %esp, %ebp");*/
   zenter();
   statemen();
   /*ol("movl %ebp, %esp");
   ol("popl %ebp");*/
   zleave();
   zret();
   cg_print(&codeg);
   cg_done(&codeg);
   /*fprintf(stderr,"dofunc:savecg=%d\n",savecg);*/
   ccg=savecg;
   cnmlst=savenmlst;
   /*fprintf(stderr,"dofunc:ccg=%d\n",ccg);*/
}
func _zret()
{
   ol("ret");
}
func statemen()
{
   /*fprintf(stderr,"statemen():ccg=%d\n",ccg);*/
   blanks();
   if((!ch())&&iseof)return 0;
   else
   if(amatch("var",3))dolocvar();
   else if(match(tlcomp/*"{"*/))compound();
   else if(amatch("if",2))
     {
       doif();
       laatst=stif;
     }
   else if(amatch("while",5))
     {
     dowhile();
     laatst=stwhile;
     }
   else if(amatch("for",3))
     {
     dofor();
```

```
      lastst=stfor;
      }
   else if(amatch("do",2))
      {
      dodo();
      lastst=stdo;
      }
   else if(amatch("return",6))
      {
      doreturn();
      ns();
      lastst=streturn;
      }
   else if(amatch("break",5))
      {
      dobreak();
      ns();
      lastst=stbreak;
      }
   else if(amatch("continue",8))
      {
      docont();
      ns();
      lastst=stcont;
      }
   else if(match(";"));
   else
      {
      expressi();
      ns();
      lastst=stexp;
      }
   return lastst;
}
func doreturn()
{
   if(!endst())expressi();
   zleave();
   /*ol("movl %ebp, %esp");
   ol("popl %ebp");*/
   zret();
}
func compound()
{
   ++ncmp;
   while(!match(trcomp/*"}"*/))
   {
      statemen();
      if(iseof)break;
```

```
   }
   --ncmp;
}
func modstk(newsp:int)
{
  var int:k;
  k=newsp-Zsp;
  if(!k)return newsp;
  cmodstk(k);
  return newsp;
  if(k>0){
  ot("addl $");
  outdec(k);
  outasm(", %esp");
  nl();
  return newsp;
  }
  else
  {
    ot("subl $");
    outdec(-k);
    outasm(", %esp");
    nl();
    return newsp;
  }
}
func dolocvar()
{
  var [NAMESIZE]char:sname;
  var int:p1;
  var int:typ,istyp;
  var *ssym:idx;
  var int:k;
  var *ssymlist:lst,lptr;
  var **ssymlist:lpom;
  lst=lptr=0;
  lpom=&lst;
  istyp=0;
  if(cbtype())
  {
    /*fprintf(stderr,"line=%s\n",line);
    fprintf(stderr,"can be type\n");*/
    istyp=1;
    typ=gettypen();
    if(!match(":"))error("':' expected");
    k=gettsize(typ);
    comment();
    outstr("size:");
    outdec(k);
```

```
  nl();
  if(k&3)k=k+4-(k&3);
  comment();
  outdec(k);
  nl();
}
blanks();
if(!an(ch()))error("expected name of variable");
/*p1=locptr;*/
while(symname(sname))
{
  chkmem((*lpom)=calloc(1,sizeof(ssymlist)));
  strcp((*lpom)->sym.name,sname);
  lpom=&(*lpom)->next;
  (*lpom)=0;
  if(istyp)
  {if(match(";"))break;}
  else
  {if(match(":"))break;}
  if(endst())break;
  if(!match(","))
  {
    error("',' or ':'|';' expected");
    junk();
    break;
  }
}
if(!istyp)
{
  typ=gettypen();
  ns();
  k=gettsize(typ);
  comment();
  outstr("size:");
  outdec(k);
  nl();
  if(k&3)k=k+4-(k&3);
  comment();
  outdec(k);
  nl();
}
for(lptr=lst;lptr;lptr=lptr->next)
{
  idx=findloc(lptr->sym.name);
  /*fprintf(stderr,"adding %s\n",lptr->sym.name);*/
  if(idx)
  {
    error("local multidef");
  }
```

```
    addloc(lptr->sym.name,S_VARL,1,Zsp-k,typ);
    Zsp=modstk(Zsp-k);
  }
  /*  while(symname(sname))
  {
    idx=findloc(sname);
    if(idx){
    error("local multidef");
    }
    addloc(sname,S_VARL,1,Zsp-k,typ);
    Zsp=modstk(Zsp-k);
    if(match(";"))break;
    if(endst())break;
    if(!match(","))
    {
      error("',' or ';' expected");
      junk();
      break;
    }
    }*/
  delsymlist(lst);
}
func doif()
{
  /*var int:flev;*/
  var **ssymlist:flev;
  var int:fsp;
  var int:flab1;
  var int:flab2;
  flev=locsymtab.front;
  /*flev=locptr;*/
  fsp=Zsp;
  flab1=getlabel();
  test(flab1);
  statemen();
  Zsp=modstk(fsp);
  /*locptr=flev;*/
  ssymtabcut(&locsymtab,flev);
  if(!amatch("else",4)){
  /*printlab(flab1);
  col();
  nl();*/
  clab(flab1);
  return ;
  }
  jump(flab2=getlabel());
  /*printlab(flab1);
  col();
  nl();*/
```

```
    clab(flab1);
    statemen();
    Zsp=modstk(fsp);
    /*locptr=flev;*/
    ssymtabcut(&locsymtab,flev);
    /*printlab(flab2);
    col();
    nl();
    */
    clab(flab2);
}
func test(label:int)
{
    needbrac(tlarg/*"("*/);
    expressi();
    needbrac(trarg/*")"*/);
    testjump(label);
}
func dobreak()
{
    var int:ptr;
    if(!wqptr)return ;
    ptr=readwhil();
    modstk(wqsp[ptr]);
    jump(wqlab[ptr]);
}
func docont()
{
    var int:ptr;
    if(!wqptr)return ;
    ptr=readwhil();
    modstk(wqsp[ptr]);
    jump(wqloop[ptr]);
}
func dofor()
{
    var **ssymlist:thesym;
    var int:thesp;
    var int:theloop,thecont;
    var int:thelab;
    var int:tl;
    /*thesym=locptr;*/
    thesym=locsymtab.front;
    thesp=Zsp;
    theloop=getlabel();
    thelab=getlabel();
    thecont=getlabel();
    addwhile(thesym,thesp,thecont,thelab);
    needbrac(tlarg/*"("*/);
```

```
        expressi();/* i=0 */
        ns();
        /*printlab(theloop);
        col();
        nl();*/
        clab(theloop);
        expressi();/* i<N */
        testjump(thelab);
        ns();
        /*fprintf(stderr,"dofor:ccg=%d\n",ccg);*/
        var *scodegen:savecg;
        savecg=ccg;
        /*fprintf(stderr,"dofor:savecg=%d\n",savecg);*/
        var scodegen:cg1;
        cg_init(&cg1);
        ccg=&cg1;
        tl=getlitst();
        /*printlab(thecont);col();nl();*/
        clab(thecont);
        blanks();
        if(!streq(line+lptr,trarg)/*ch()!=')'*/)
        expressi();/* i++ */
        var scodegen:cg2;
        cg_init(&cg2);
        ccg=&cg2;
        getlitst();
        needbrac(trarg/*")"*/);
        statemen();
        Zsp=modstk(thesp);
        /*cg_print(&cg2);*/
        cg_transfer(&cg2,savecg);
        cg_done(&cg2);
        /*cg_print(&cg1);*/
        cg_transfer(&cg1,savecg);
        cg_done(&cg1);
        ccg=savecg;
        dumpltst(tl);
        jump(theloop);
        /*printlab(thelab);
        col();
        nl();*/
        clab(thelab);
        /*locptr=thesym;*/
        ssymtabcut(&locsymtab,thesym);
        delwhile();
}
func dodo()
{
    var *ssymlist:thesym;
```

```
    var int:thesp;
    var int:theloop;
    var int:thelab;
    /*thesym=locptr;*/
    thesym=locsymtab.front;
    thesp=Zsp;
    theloop=getlabel();
    thelab=getlabel();
    addwhile(thesym,thesp,theloop,thelab);
    /*printlab(theloop);
    col();
    nl();*/
    clab(theloop);
    statemen();
    Zsp=modstk(thesp);
    if(!amatch("while",5)){
    error("'while' expected");
    }
    needbrac(tlarg/*"("*/);
    expressi();
    /*ol("testl %eax, %eax");
    ot("jne");
    tab();
    printlab(theloop);
    nl();*/
    testnejump(theloop);
    /*printlab(thelab);
    col();
    nl();*/
    clab(thelab);
    needbrac(trarg/*")"*/);
    ns();
    /*locptr=thesym;*/
    ssymtabcut(&locsymtab,thesym);
    delwhile();
}
func dowhile()
{
    var **ssymlist:thesym;
    var int:thesp;
    var int:theloop;
    var int:thelab;
    /*thesym=locptr;*/
    thesym=locsymtab.front;
    thesp=Zsp;
    theloop=getlabel();
    thelab=getlabel();
    addwhile(thesym,thesp,theloop,thelab);
    /*printlab(theloop);
```

```
  col();
  nl();*/
  clab(theloop);
  test(thelab);
  statemen();
  Zsp=modstk(thesp);
  jump(theloop);
  /*printlab(thelab);
  col();
  nl();*/
  clab(thelab);
  /*locptr=thesym;*/
  ssymtabcut(&locsymtab,thesym);
  delwhile();
}
func addwhile(sym:int,sp:int,loop:int,lab:int)
{
  if(wqptr>=WQMAX){
  error("too many nested loops");
  return ;
  }
  wqsym[wqptr]=sym;
  wqsp[wqptr]=sp;
  wqloop[wqptr]=loop;
  wqlab[wqptr]=lab;
  wqptr++;
}
func readwhil()
{
  if(!wqptr){
  error("no active loops");
  return 0;
  }
  return wqptr-1;
}
func delwhile()
{
  if(wqptr>0)wqptr--;
}
func _zpush()
{
  ol("pushl %eax");
  Zsp=Zsp-4;
}
func _mult()
{
  ol("imull %edx");
}
func _zpop()
```

```
{
  ol("popl %edx");
  Zsp=Zsp+4;
}
func _zmod()
{
  div();
  ol("movl %edx, %eax");
}
func _div()
{
  ol("xchgl %eax, %edx");
  ol("movl %edx, %ecx");
  ol("cltd");
  ol("idivl %ecx");
}
func _increg(k:int)
{
  if(!k)return ;
  if(k<3)while(k--)ol("incl %eax");
  else
  {
    ot("addl $");
    outdec(k);
    outstr(", %eax");
    nl();
  }
}
func _decreg(k:int)
{
  if(!k)return ;
  if(k<3)while(k--)ol("decl %eax");
  else
  {
    ot("subl $");
    outdec(k);
    outstr(", %eax");
    nl();
  }
}
func _mulreg(k:int,s:*char)
{
  var int:l;
  if(k==1)return ;
  else
  if(k==0){
    ot("xorl ");
    outstr(s);
    outstr(", ");
```

```
      outstr(s);
      nl();
    }
    else
      {
      l=1;
      while(l<15)if(k==(1<<l)){
        ot("sall $");
        outdec(l);
        outstr(", ");
        outstr(s);
        nl();
        return ;
      }
      else
        l++;
      }
    ot("imull $");
    outdec(k);
    outstr(", ");
    outstr(s);
    nl();
}
func _divconst(k:int)
{
    var int:l;
    if(k==1)return ;
    if(!k){
    error("division by zero");
    return ;
    }
    l=1;
    while(l<15)if(k==(1<<l)){
    ot("sarl $");
    outdec(l);
    outstr(", %eax");
    nl();
    return ;
    }
    else
    l++;
    ol("cltd");
    ot("divl $");
    outdec(k);
    nl();
}
func _zadd()
{
    ol("addl %edx, %eax");
```

```
}
func _zsub()
{
  ol("subl %eax, %edx");
  ol("movl %edx, %eax");
}
func _neg()
{
  ol("negl %eax");
}
func _zeq()
{
  ot("cmpl");
  ot("%eax, %edx");
  nl();
  ot("sete");
  ot("%al");
  nl();
  ot("movzbl");
  ot("%al, %eax");
  nl();
}
func _zne()
{
  ot("cmpl");
  ot("%eax, %edx");
  nl();
  ot("setne");
  ot("%al");
  nl();
  ot("movzbl");
  ot("%al, %eax");
  nl();
}
func _zge()
{
  ot("cmpl");
  ot("%eax, %edx");
  nl();
  ot("setge");
  ot("%al");
  nl();
  ot("movzbl");
  ot("%al, %eax");
  nl();
}
func _uge()
{
  ot("cmpl");
```

```
    ot("%eax, %edx");
    nl();
    ot("setae");
    ot("%al");
    nl();
    ot("movzbl");
    ot("%al, %eax");
    nl();
}
func _ule()
{
    ot("cmpl");
    ot("%eax, %edx");
    nl();
    ot("setbe");
    ot("%al");
    nl();
    ot("movzbl");
    ot("%al, %eax");
    nl();
}
func _zle()
{
    ot("cmpl");
    ot("%eax, %edx");
    nl();
    ot("setle");
    ot("%al");
    nl();
    ot("movzbl");
    ot("%al, %eax");
    nl();
}
func _ult()
{
    ot("cmpl");
    ot("%eax, %edx");
    nl();
    ot("setb");
    ot("%al");
    nl();
    ot("movzbl");
    ot("%al, %eax");
    nl();
}
func _zlt()
{
    ot("cmpl");
    ot("%eax, %edx");
```

```
  nl();
  ot("setl");
  ot("%al");
  nl();
  ot("movzbl");
  ot("%al, %eax");
  nl();
}
func _ugt()
{
  ot("cmpl");
  ot("%eax, %edx");
  nl();
  ot("seta");
  ot("%al");
  nl();
  ot("movzbl");
  ot("%al, %eax");
  nl();
}
func _zgt()
{
  ot("cmpl");
  ot("%eax, %edx");
  nl();
  ot("setg");
  ot("%al");
  nl();
  ot("movzbl");
  ot("%al, %eax");
  nl();
}
func _zor()
{
  ol("orl %edx, %eax");
}
func _zxor()
{
  ol("xorl %edx, %eax");
}
func _zand()
{
  ol("andl %edx, %eax");
}
func _asr()
{
  ol("movl %eax, %ecx");
  ol("movl %edx, %eax");
  ol("sarl %cl, %eax");
```

```
}
func _asl()
{
  ol("movl %eax, %ecx");
  ol("movl %edx, %eax");
  ol("sall %cl, %eax");
}
func _lnot()
{
  ol("testl %eax,%eax");
  ol("sete %al");
  ol("movzbl %al, %eax");
}
func _bnot()
{
  ol("notl %eax");
}
func _testjump(label:int)
{
  ol("testl %eax, %eax");
  ot("je");
  tab();
  printlab(label);
  nl();
}
func _testnejump(label:int)
{
  ol("testl %eax, %eax");
  ot("jne");
  tab();
  printlab(label);
  nl();
}
func _jump(label:int)
{
  ot("jmp ");
  printlab(label);
  nl();
}
func _zcall(sname:*char)
{
  ot("call ");
  outname(sname);
  nl();
}
func printlab(label:int)
{
  outasm(".L");
  outdec(label);
```

```
}
func indirect(lval:*int)
{
}

func pretree(node:*enode,ofil:*int)
{
  if(!node)
  return;
  fprintf(ofil,"(");
  pretree(node->l,ofil);
  if(node->op==OP_LEAF)
  {
    if(node->leaf.vid==L_NUM)
    fprintf(ofil,"%d",node->leaf.val);
    else if(node->leaf.vid==L_STR)
    fprintf(ofil,"\"\"",litq+node->leaf.val);
    else if(node->leaf.vid==L_ID)
    fprintf(ofil,"%s",node->name);
    else
    error("unknown leaf");
  }
  else if(node->op==OP_LIST)
  {
    fprintf(ofil,"[]");
  }
  else if(node->op==OP_FUNC)
  {
    fprintf(ofil,"$");
  }
  else if(node->op==OP_COMMA)
  {
    fprintf(ofil,",");
  }
  else if(node->op==OP_COND)
  {
    fprintf(ofil,"?");
  }
  else if(node->op==OP_ASSIGN)
  {
    fprintf(ofil,"=");
  }
  else if(node->op==OP_EQ)
  {
    fprintf(ofil,"==");
  }
  else if(node->op==OP_NEQ)
  {
    fprintf(ofil,"!=");
```

```
}
else if(node->op==OP_GT)
{
  fprintf(ofil,">");
}
else if(node->op==OP_LT)
{
  fprintf(ofil,"<");
}
else if(node->op==OP_GE)
{
  fprintf(ofil,">=");
}
else if(node->op==OP_LE)
{
  fprintf(ofil,"<=");
}
else if(node->op==OP_SHL)
{
  fprintf(ofil,"<<");
}
else if(node->op==OP_SHR)
{
  fprintf(ofil,">>");
}
else if(node->op==OP_BAND)
{
  fprintf(ofil,"&");
}
else if(node->op==OP_BOR)
{
  fprintf(ofil,"|");
}
else if(node->op==OP_BXOR)
{
  fprintf(ofil,"^");
}
else if(node->op==OP_LAND)
{
  fprintf(ofil,"&&");
}
else if(node->op==OP_LOR)
{
  fprintf(ofil,"||");
}
else if(node->op==OP_PLUS)
{
  fprintf(ofil,"+");
}
```

```c
else if(node->op==OP_MINUS)
{
  fprintf(ofil,"-");
}
else if(node->op==OP_MUL)
{
  fprintf(ofil,"*");
}
else if(node->op==OP_DIV)
{
  fprintf(ofil,"/");
}
else if(node->op==OP_REM)
{
  fprintf(ofil,"%");
}
else if(node->op==OP_1PP)
{
  fprintf(ofil,"++@");
}
else if(node->op==OP_1MM)
{
  fprintf(ofil,"--@");
}
else if(node->op==OP_UMINUS)
{
  fprintf(ofil,"u-");
}
else if(node->op==OP_2PP)
{
  fprintf(ofil,"@++");
}
else if(node->op==OP_2MM)
{
  fprintf(ofil,"@--");
}
else if(node->op==OP_STAR)
{
  fprintf(ofil,"*@");
}
else if(node->op==OP_ADDR)
{
  fprintf(ofil,"&@");
}
else if(node->op==OP_BNOT)
{
  fprintf(ofil,"~");
}
else if(node->op==OP_LNOT)
```

```
  {
    fprintf(ofil,"!");
  }
  else if(node->op==OP_DOT)
  {
    fprintf(ofil,".%s",node->name);
  }
  else error("unknown op");
  pretree(node->r,ofil);
  if(node->op==OP_COND)
  {
    fprintf(ofil,":");
    pretree(node->third,ofil);
  }
  fprintf(ofil,")");
}

func cttype(node:*enode)
{
  if(!node)return T_INT;
  if(node->op==OP_LEAF)
  {
    if(node->leaf.vid==L_NUM)
    return T_INT;
    else if(node->leaf.vid==L_ID)
    return node->leaf.idx->type;
    else if(node->leaf.vid==L_STR)
    return T_CHARP;
  }
  else if(node->op==OP_STAR)
  {
    var int:t;
    t=cttype(node->r);
    if(typtab[t].sort!=V_PTR&&typtab[t].sort!=V_PTR)
    {error("star of non-ptr");return T_INT;}
    return(typtab[t].type);
  }
  else if(node->op==OP_ADDR)
  {
    var int:t;
    t=cttype(node->r);
    if(typtab[t].sort==V_ARR)
    {error("extre '&' for array");return t;}
    return getptrty(t);
  }
  else if(node->op==OP_PLUS)
  {
    var int:t1,t2;
  }
```

```
    else if(node->op==OP_MINUS)
    {
    }
    else error("don't know the type");
}
func treetocode(node:*enode,lval:*elval)
{
  if(!node)
  return 0;
  if(node->op==OP_LEAF)
    {
      if(node->leaf.vid==L_NUM)
        {
          lval->sort=L_NUM;
          lval->idx=0;
          lval->offset=0;
          lval->typ=T_INT;
          lval->val=node->leaf.val;
          return 1;
        }
      else if(node->leaf.vid==L_ID)
        {
          lval->sort=L_ID;
          lval->idx=node->leaf.idx;
          lval->offset=0;
          lval->typ=node->leaf.idx->type;
          strcp(lval->name,node->name);
          return 1;
        }
      else if(node->leaf.vid==L_STR)
        {
          lval->sort=L_STR;
          lval->idx=0;
          lval->offset=node->leaf.val;
          lval->typ=T_CHARP;
          return 1;
        }
      else
      error("how to code a leaf?");
    }
  else if(node->op==OP_COMMA)
  return ct_COMMA(node,lval);
  else if(node->op==OP_LOR)
  return ct_LOR(node,lval);
  else if(node->op==OP_LAND)
  return ct_LAND(node,lval);
  else if(node->op==OP_BOR)
  return ct_BOR(node,lval);
  else if(node->op==OP_BXOR)
```

```
return ct_BXOR(node,lval);
else if(node->op==OP_BAND)
return ct_BAND(node,lval);
else if(node->op==OP_EQ)
return ct_EQ(node,lval);
else if(node->op==OP_NEQ)
return ct_NEQ(node,lval);
else if(node->op==OP_GT)
return ct_GT(node,lval);
else if(node->op==OP_LT)
return ct_LT(node,lval);
else if(node->op==OP_GE)
return ct_GE(node,lval);
else if(node->op==OP_LE)
return ct_LE(node,lval);
else if(node->op==OP_SHL)
return ct_SHL(node,lval);
else if(node->op==OP_SHR)
return ct_SHR(node,lval);
else if(node->op==OP_PLUS)
return ct_PLUS(node,lval);
else if(node->op==OP_MINUS)
return ct_MINUS(node,lval);
else if(node->op==OP_ASSIGN)
return ct_ASSIGN(node,lval);
else if(node->op==OP_MUL)
return ct_MUL(node,lval);
else if(node->op==OP_DIV)
return ct_DIV(node,lval);
else if(node->op==OP_REM)
return ct_REM(node,lval);
else if(node->op==OP_1PP)
return ct_1PP(node,lval);
else if(node->op==OP_1MM)
return ct_1MM(node,lval);
else if(node->op==OP_2PP)
return ct_2PP(node,lval);
else if(node->op==OP_2MM)
return ct_2MM(node,lval);
else if(node->op==OP_UMINUS)
return ct_UMINUS(node,lval);
else if(node->op==OP_STAR)
return ct_STAR(node,lval);
else if(node->op==OP_ADDR)
return ct_ADDR(node,lval);
else if(node->op==OP_LNOT)
return ct_LNOT(node,lval);
else if(node->op==OP_BNOT)
return ct_BNOT(node,lval);
```

```
    else if(node->op==OP_FUNC)
    return ct_FUNC(node,lval);
    else if(node->op==OP_DOT)
    return ct_DOT(node,lval);
    else error("to be implemented");
}
func ct_COMMA(node:*enode,lval:*elval)
{
    var *enode:cnode;
    var int:k;
    var elval:lval1;
    for(cnode=node;cnode;cnode=cnode->r)
    {
        k=treetocode(cnode->l,&lval1);
        if(k)rvalue(&lval1);
        if(cnode->r&&cnode->r->op!=OP_COMMA)
            return treetocode(cnode->r,lval);
    }
}
func ct_ASSIGN(node:*enode,lval:*elval)
{
    var int:k;
    var elval:lval1,lval2;
    k=treetocode(node->l,&lval1);
    if(!k){needlval();return 0;}
    if(lval1.sort==L_POI)
    {
        zpush();
        lval1.sort=L_SP;
    }
    if(treetocode(node->r,&lval2))rvalue(&lval2);
    lval->sort=L_ONREG;
    lval->idx=0;
    lval->offset=0;
    lval->typ=lval2.typ;
    store(&lval1);
    return 0;
}
func ct_LOR(node:*enode,lval:*elval)
{
    var elval:lval1,lval2;
    var *enode:cnode;
    var int:onelab,exitlab;
    if(treetocode(node->l,&lval1))rvalue(&lval1);
    onelab=getlabel();
    exitlab=getlabel();
    testnejump(onelab);
    cnode=node->r;
    while(cnode->op==OP_LOR)
```

```
  {
    if(treetocode(cnode->l,&lval1))rvalue(&lval1);
    testnejump(onelab);
    cnode=cnode->r;
  }
  if(treetocode(cnode,&lval2))rvalue(&lval2);
  testnejump(onelab);
  /*ol("xorl %eax, %eax");*/
  zldn(0);
  jump(exitlab);
  /*printlab(onelab);col();nl();*/
  clab(onelab);
  /*ol("movl $1, %eax");*/
  zldn(1);
  /*printlab(exitlab);col();nl();*/
  clab(exitlab);
  lval->sort=L_ONREG;
  lval->idx=0;
  lval->offset=0;
  lval->typ=T_INT;
  return 0;
}
func ct_LAND(node:*enode,lval:*elval)
{
  var elval:lval1,lval2;
  var *enode:cnode;
  var int:zerolab,exitlab;
  if(treetocode(node->l,&lval1))rvalue(&lval1);
  zerolab=getlabel();
  exitlab=getlabel();
  testjump(zerolab);
  cnode=node->r;
  while(cnode->op==OP_LAND)
  {
    if(treetocode(cnode->l,&lval1))rvalue(&lval1);
    testjump(zerolab);
    cnode=cnode->r;
  }
  if(treetocode(cnode,&lval2))rvalue(&lval2);
  testjump(zerolab);
  /*ol("movl $1, %eax");*/
  zldn(1);
  jump(exitlab);
  /*printlab(zerolab);col();nl();*/
  clab(zerolab);
  /*ol("xorl %eax, %eax");*/
  zldn(0);
  /*printlab(exitlab);col();nl();*/
  clab(exitlab);
```

```
  lval->sort=L_ONREG;
  lval->idx=0;
  lval->offset=0;
  lval->typ=T_INT;
  return 0;
}
func ct_BOR(node:*enode,lval:*elval)
{
  var elval:lval1,lval2;
  if(treetocode(node->l,&lval1))rvalue(&lval1);
  zpush();
  if(treetocode(node->r,&lval2))rvalue(&lval2);
  zpop();
  zor();
  lval->sort=L_ONREG;
  lval->idx=0;
  lval->offset=0;
  lval->typ=T_INT;
  return 0;
}
func ct_BXOR(node:*enode,lval:*elval)
{
  var elval:lval1,lval2;
  if(treetocode(node->l,&lval1))rvalue(&lval1);
  zpush();
  if(treetocode(node->r,&lval2))rvalue(&lval2);
  zpop();
  zxor();
  lval->sort=L_ONREG;
  lval->idx=0;
  lval->offset=0;
  lval->typ=T_INT;
  return 0;
}
func ct_BAND(node:*enode,lval:*elval)
{
  var elval:lval1,lval2;
  if(treetocode(node->l,&lval1))rvalue(&lval1);
  zpush();
  if(treetocode(node->r,&lval2))rvalue(&lval2);
  zpop();
  zand();
  lval->sort=L_ONREG;
  lval->idx=0;
  lval->offset=0;
  lval->typ=T_INT;
  return 0;
}
func ct_EQ(node:*enode,lval:*elval)
```

```
{
  var elval:lval1,lval2;
  if(treetocode(node->l,&lval1))rvalue(&lval1);
  zpush();
  if(treetocode(node->r,&lval2))rvalue(&lval2);
  zpop();
  zeq();
  lval->sort=L_ONREG;
  lval->idx=0;
  lval->offset=0;
  lval->typ=T_INT;
  return 0;
}
func ct_NEQ(node:*enode,lval:*elval)
{
  var elval:lval1,lval2;
  if(treetocode(node->l,&lval1))rvalue(&lval1);
  zpush();
  if(treetocode(node->r,&lval2))rvalue(&lval2);
  zpop();
  zne();
  lval->sort=L_ONREG;
  lval->idx=0;
  lval->offset=0;
  lval->typ=T_INT;
  return 0;
}
func ct_GT(node:*enode,lval:*elval)
{
  var elval:lval1,lval2;
  if(treetocode(node->l,&lval1))rvalue(&lval1);
  zpush();
  if(treetocode(node->r,&lval2))rvalue(&lval2);
  zpop();
  if(!issigned(lval1.typ)||!issigned(lval2.typ))
  ugt();
  else
  zgt();

  lval->sort=L_ONREG;
  lval->idx=0;
  lval->offset=0;
  lval->typ=T_INT;
  return 0;
}
func ct_LT(node:*enode,lval:*elval)
{
  var elval:lval1,lval2;
  if(treetocode(node->l,&lval1))rvalue(&lval1);
```

```
    zpush();
    if(treetocode(node->r,&lval2))rvalue(&lval2);
    zpop();
    if(!issigned(lval1.typ)||!issigned(lval2.typ))
    ult();
    else
    zlt();

    lval->sort=L_ONREG;
    lval->idx=0;
    lval->offset=0;
    lval->typ=T_INT;
    return 0;
}
func ct_GE(node:*enode,lval:*elval)
{
    var elval:lval1,lval2;
    if(treetocode(node->l,&lval1))rvalue(&lval1);
    zpush();
    if(treetocode(node->r,&lval2))rvalue(&lval2);
    zpop();
    if(!issigned(lval1.typ)||!issigned(lval2.typ))
    uge();
    else
    zge();

    lval->sort=L_ONREG;
    lval->idx=0;
    lval->offset=0;
    lval->typ=T_INT;
    return 0;
}
func ct_LE(node:*enode,lval:*elval)
{
    var elval:lval1,lval2;
    if(treetocode(node->l,&lval1))rvalue(&lval1);
    zpush();
    if(treetocode(node->r,&lval2))rvalue(&lval2);
    zpop();
    if(!issigned(lval1.typ)||!issigned(lval2.typ))
    ule();
    else
    zle();

    lval->sort=L_ONREG;
    lval->idx=0;
    lval->offset=0;
    lval->typ=T_INT;
    return 0;
```

```
}
func ct_SHL(node:*enode,lval:*elval)
{
  var elval:lval1,lval2;
  if(treetocode(node->l,&lval1))rvalue(&lval1);
  zpush();
  if(treetocode(node->r,&lval2))rvalue(&lval2);
  zpop();
  asl();
  lval->sort=L_ONREG;
  lval->idx=0;
  lval->offset=0;
  lval->typ=T_INT;
  return 0;
}
func ct_SHR(node:*enode,lval:*elval)
{
  var elval:lval1,lval2;
  if(treetocode(node->l,&lval1))rvalue(&lval1);
  zpush();
  if(treetocode(node->r,&lval2))rvalue(&lval2);
  zpop();
  asr();
  lval->sort=L_ONREG;
  lval->idx=0;
  lval->offset=0;
  lval->typ=T_INT;
  return 0;
}
func ct_MINUS(node:*enode,lval:*elval)
{
  var elval:lval1,lval2;
  var int:isptr;
  isptr=0;
  if(treetocode(node->l,&lval1))rvalue(&lval1);
  zpush();
  if(treetocode(node->r,&lval2))rvalue(&lval2);
  zpop();
  lval->sort=L_ONREG;
  lval->idx=0;
  lval->offset=0;
  lval->typ=T_INT;
  if(typtab[lval1.typ].sort==V_PTR||
   typtab[lval1.typ].sort==V_ARR)
   {
     if(typtab[lval2.typ].sort==V_PTR||
      typtab[lval2.typ].sort==V_ARR)
      {
        zsub();
```

```
      if(typtab[lval1.typ].type!=typtab[lval2.typ].type)
      error("subtracting pointers of different types");
      divconst(gettsize(typtab[lval1.typ].type));
      isptr=0;/*?*/
      lval->typ=T_INT;/*?*/
    }
    else
    {
      mulreg(gettsize(typtab[lval1.typ].type),/*"%eax"*/RG_A);
      zsub();
      lval->typ=lval1.typ;
    }
  }
  else
  {
    if(typtab[lval2.typ].sort==V_PTR||
     typtab[lval2.typ].sort==V_ARR)
    error("subtracting a pointer");
    zsub();
  }
  return 0;/*on register*/
}
func ct_PLUS(node:*enode,lval:*elval)
{
  var elval:lval1,lval2;
  var int:isptr;
  isptr=0;
  if(treetocode(node->l,&lval1))rvalue(&lval1);
  zpush();
  if(treetocode(node->r,&lval2))rvalue(&lval2);
  lval->sort=L_ONREG;
  lval->idx=0;
  lval->offset=0;
  lval->typ=T_INT;
  if(typtab[lval1.typ].sort==V_PTR||
   typtab[lval1.typ].sort==V_ARR)
   {
     mulreg(gettsize(typtab[lval1.typ].type),/*"%eax"*/RG_A);
     isptr=1;
     lval->typ=lval1.typ;
   }
  zpop();
  if(typtab[lval2.typ].sort==V_PTR||
   typtab[lval2.typ].sort==V_ARR)
   {
     if(isptr)error("cannot add pointer to pointer");
     mulreg(gettsize(typtab[lval2.typ].type),/*"%edx"*/RG_D);
     isptr=1;/*legacy?*/
     lval->typ=lval2.typ;
```

```
  }
  zadd();
  return 0;/*on register*/
}
func ct_MUL(node:*enode,lval:*elval)
{
  var elval:lval1,lval2;
  if(treetocode(node->l,&lval1))rvalue(&lval1);
  zpush();
  if(treetocode(node->r,&lval2))rvalue(&lval2);
  zpop();
  mult();
  lval->sort=L_ONREG;
  lval->idx=0;
  lval->offset=0;
  lval->typ=T_INT;
  return 0;
}
func ct_DIV(node:*enode,lval:*elval)
{
  var elval:lval1,lval2;
  if(treetocode(node->l,&lval1))rvalue(&lval1);
  zpush();
  if(treetocode(node->r,&lval2))rvalue(&lval2);
  zpop();
  div();
  lval->sort=L_ONREG;
  lval->idx=0;
  lval->offset=0;
  lval->typ=T_INT;
  return 0;
}
func ct_REM(node:*enode,lval:*elval)
{
  var elval:lval1,lval2;
  if(treetocode(node->l,&lval1))rvalue(&lval1);
  zpush();
  if(treetocode(node->r,&lval2))rvalue(&lval2);
  zpop();
  zmod();
  lval->sort=L_ONREG;
  lval->idx=0;
  lval->offset=0;
  lval->typ=T_INT;
  return 0;
}
func ct_1PP(node:*enode,lval:*elval)
{
  var int:k,sz;
```

```
  k=treetocode(node->r,lval);
  if(!k){needlval();return 0;}
  if(lval->sort==L_POI)
  {
    zpush();
    rvalue(lval);
    lval->sort=L_SP;
  }
  else rvalue(lval);
  if(typtab[lval->typ].sort==V_PTR)
  sz=gettsize(typtab[lval->typ].type);
  else sz=1;
  increg(sz);
  store(lval);
  lval->sort=L_ONREG;
  lval->idx=0;
  lval->offset=0;
  return 0;
}
func ct_1MM(node:*enode,lval:*elval)
{
  var int:k,sz;
  k=treetocode(node->r,lval);
  if(!k){needlval();return 0;}
  if(lval->sort==L_POI)
  {
    zpush();
    rvalue(lval);
    lval->sort=L_SP;
  }
  else rvalue(lval);
  if(typtab[lval->typ].sort==V_PTR)
  sz=gettsize(typtab[lval->typ].type);
  else sz=1;
  decreg(sz);
  store(lval);
  lval->sort=L_ONREG;
  lval->idx=0;
  lval->offset=0;
  return 0;
}
func ct_2PP(node:*enode,lval:*elval)
{
  var int:k,sz;
  k=treetocode(node->r,lval);
  if(!k){needlval();return 0;}
  if(lval->sort==L_POI)
  {
    zpush();
```

```
      rvalue(lval);
      lval->sort=L_SP;
   }
   else rvalue(lval);
   if(typtab[lval->typ].sort==V_PTR)
   sz=gettsize(typtab[lval->typ].type);
   else sz=1;
   increg(sz);
   store(lval);
   decreg(sz);
   lval->sort=L_ONREG;
   lval->idx=0;
   lval->offset=0;
   return 0;
}
func ct_2MM(node:*enode,lval:*elval)
{
   var int:k,sz;
   k=treetocode(node->r,lval);
   if(!k){needlval();return 0;}
   if(lval->sort==L_POI)
   {
      zpush();
      rvalue(lval);
      lval->sort=L_SP;
   }
   else rvalue(lval);
   if(typtab[lval->typ].sort==V_PTR)
   sz=gettsize(typtab[lval->typ].type);
   else sz=1;
   decreg(sz);
   store(lval);
   increg(sz);
   lval->sort=L_ONREG;
   lval->idx=0;
   lval->offset=0;
   return 0;
}
func ct_UMINUS(node:*enode,lval:*elval)
{
   var int:k;
   k=treetocode(node->r,lval);
   if(k)rvalue(lval);
   neg();
   lval->sort=L_ONREG;
   lval->idx=0;
   lval->offset=0;
   return 0;
}
```

```
func ct_STAR(node:*enode,lval:*elval)
{
  var int:k;
  k=treetocode(node->r,lval);
  if(lval->sort==L_ID||lval->sort!=L_STR)
  {
    if(typtab[lval->typ].sort!=V_ARR&&typtab[lval->typ].sort!=V_PTR)
    {error("need a pointer");return k;}
    if(k)rvalue(lval);
    lval->sort=L_POI;
    lval->typ=typtab[lval->typ].type;
    lval->offset=0;
    return 1;
  }
  else if(lval->sort==L_STR)
  {
    if(k)rvalue(lval);
    lval->sort=L_POI;
    lval->typ=T_CHAR;
    lval->offset=0;
    return 1;
  }
  else{error("error with *pointe");return 0;}
}
func ct_ADDRDIR(node:*enode,lval:*elval)
{
  var int:k;
  k=treetocode(node,lval);
  if(lval->sort==L_ID&&lval->idx)
  {
    if(lval->idx->sort==S_VARG)
    {
      /*ot("movl $");
      outname(lval->idx->name);
      if(lval->offset)
      {outasm("+");outdec(lval->offset);}
      outasm(", %eax");
      nl();*/
      zlda(lval->idx->name,lval->offset);
    }
    else if(lval->idx->sort==S_VARL)
    {
      /*ot("leal ");
      outdec(lval->idx->offset+lval->offset);
      outasm("(%ebp), %eax");
      nl();*/
      zlea(lval->idx->offset+lval->offset);
    }
    else if(lval->idx->sort==S_FUNC)
```

```
      {
        error("address of function: to be implemented");
      }
      else error("error taking address");
    }
    else if(lval->sort==L_NUM)
    error("how to take the address of a number?");
    else if(lval->sort==L_STR)
    {
      loadlita(lval);
    }
    else if(lval->sort==L_POI)
    {
      if(lval->offset)
      {
        increg(lval->offset);
        lval->offset=0;
      }
    }
    else error("error after &");
    lval->typ=getptrty(lval->typ);
    return 0;
}
func ct_ADDR(node:*enode,lval:*elval)
{
    var int:k;
    return ct_ADDRDIR(node->r,lval);
    k=treetocode(node->r,lval);
    if(lval->sort==L_ID&&lval->idx)
    {
      if(lval->idx->sort==S_VARG)
      {
        /*ot("movl $");
        outname(lval->idx->name);
        if(lval->offset)
        {outasm("+");outdec(lval->offset);}
        outasm(", %eax");
        nl();*/
        zlda(lval->idx->name,lval->offset);
      }
      else if(lval->idx->sort==S_VARL)
      {
        /*ot("leal ");
        outdec(lval->idx->offset+lval->offset);
        outasm("(%ebp), %eax");
        nl();*/
        zlea(lval->idx->offset+lval->offset);
      }
      else if(lval->idx->sort==S_FUNC)
```

```
    {
      error("address of function: to be implemented");
    }
    else error("error taking address");
  }
  else if(lval->sort==L_NUM)
  error("how to take the address of a number?");
  else if(lval->sort==L_STR)
  {
    loadlita(lval);
  }
  else if(lval->sort==L_POI)
  {
    if(lval->offset)
    {
      increg(lval->offset);
      lval->offset=0;
    }
  }
  else error("error after &");
  lval->typ=getptrty(lval->typ);
  return 0;
}
func ct_LNOT(node:*enode,lval:*elval)
{
  var int:k;
  k=treetocode(node->r,lval);
  if(k)rvalue(lval);
  lnot();
  lval->sort=L_ONREG;
  lval->idx=0;
  lval->offset=0;
  lval->typ=T_INT;/*logical*/
  return 0;
}
func ct_BNOT(node:*enode,lval:*elval)
{
  var int:k;
  k=treetocode(node->r,lval);
  if(k)rvalue(lval);
  bnot();
  lval->sort=L_ONREG;
  lval->idx=0;
  lval->offset=0;
  return 0;
}
func ct_FUNC(node:*enode,lval:*elval)
{
  var int:k;
```

```
var elval:lval1,lval2;
var *enode:l,r;
l=node->l;
if(l&&((l->op==OP_LEAF&&l->leaf.vid==L_ID
 &&l->leaf.idx
 &&l->leaf.idx->sort==S_FUNC)
    ))
{
  var int:nargs;
  r=node->r;
  nargs=0;
  while(r)
  {
    k=treetocode(r->l,&lval2);
    if(k)rvalue(&lval2);
    zpush();
    nargs=nargs+WORDSIZE;
    r=r->r;
  }
  zcall(l->leaf.idx->name);
  Zsp=modstk(Zsp+nargs);
}
else if(l&&l->op==OP_DOT)
{
  /*error("calling of method: to be constructed");*/
  var int:nargs;
  r=node->r;
  nargs=0;
  while(r)
  {
    k=treetocode(r->l,&lval2);
    if(k)rvalue(&lval2);
    zpush();
    nargs=nargs+WORDSIZE;
    r=r->r;
  }
  /*zcall(l->leaf.idx->name);*/
  /*fprintf(stderr,"tree:");*/
  /*pretree(l->l,stderr);fprintf(stderr,"\n");*/
  k=ct_ADDRDIR(l->l,&lval2);
  /*fprintf(stderr,"lval2.sort=%d,offset=%d,typ=%d\n",
  lval2.sort,lval2.offset,lval2.typ);*/
  if(typtab[lval2.typ].sort!=V_PTR)
  {
    error("should be pointer...");
    fprintf(stderr,"typtab[lval2.typ]=%d\n",
        typtab[lval2.typ].sort);
  }
  if(typtab[typtab[lval2.typ].type].sort!=V_STR)
```

```
    {
      error("methods are for structures...\n");
    }
    if(k)rvalue(&lval2);
    zpush();
    nargs=nargs+WORDSIZE;
    var [NAMESIZE]char:methodname;
    strcp(methodname,typtab[typtab[lval2.typ].type].name);
    strcat(strcat(methodname,"."),l->name);/* DANGEROUS!!! FixMe! */
    cnmlst->addm(methodname);
    zcall(methodname);
    Zsp=modstk(Zsp+nargs);
  }
  else
  {
    error("function by ptr needs to be implemented");
    fprintf(stderr,
        "l->op=%d,l->leaf.vid=%d\n",
        l->op,l->leaf.vid);
  }
  lval->sort=L_ONREG;
  lval->idx=0;
  lval->offset=0;
  lval->typ=T_INT;/*the result*/
  return 0;
}
func ct_DOT(node:*enode,lval:*elval)
{
  var int:k,i;
  /*var elval:lval1,lval2;*/
  k=treetocode(node->l,lval);
  if(!k)
  {error("bizarre to '.' something on register");return k;}
  if(typtab[lval->typ].sort!=V_STR)
  {error("'.' is for structures");return k;}
  i=findfiel(node->name,lval->typ);
  if(!i)
  {error("no such field");prstrct(lval->typ);return k;}
  lval->typ=fieldtab[i].type;
  lval->offset=lval->offset+fieldtab[i].offset;
  return k;
}
func loadlita(lval:*elval)
{
  /*ot("movl $");
  printlab(stlab);
  outstr("+");
  outdec(lval->offset);
  outstr(", %eax");
```

```
   nl();*/
   cloadlita(lval->offset);
}
func store(lval:*elval)
{
   if(lval->sort==L_ID&&lval->idx)
   {
     if(lval->idx->sort==S_VARG)
     {
       if(lval->typ==T_INT||
        typtab[lval->typ].sort==V_PTR)
       /*ot("movl %eax, ");*/
       zstow(lval->idx->name,lval->offset);
       else if(lval->typ==T_CHAR)
       /*ot("movb %al, ");*/
       zstob(lval->idx->name,lval->offset);
       else error("error in global storing");
       /*outname(lval->idx->name);
       if(lval->offset)
       {outstr("+");outdec(lval->offset);}
       nl();*/
     }
     else if(lval->idx->sort==S_VARL)
     {
       if(lval->typ==T_INT||
        typtab[lval->typ].sort==V_PTR)
       /*ot("movl %eax, ");*/
       zstlw(lval->idx->offset+lval->offset);
       else if(lval->typ==T_CHAR)
       /*ot("movb %al, ");*/
       zstlb(lval->idx->offset+lval->offset);
       else error("error in local storing");
       /*outdec(lval->idx->offset+lval->offset);
       outasm("(%ebp)");
       nl();*/
     }
     else error("don't know how to store");
   }
   else if(lval->sort==L_SP)
   {
     zpop();
     if(typtab[lval->typ].size==BYTESIZE)/*ot("movb %al, ");*/
     zstob2(lval->offset);
     else if(typtab[lval->typ].size==WORDSIZE)/*ot("movl %eax, ");*/
     zstow2(lval->offset);
     else error("error storing object if strange size");
     /*outdec(lval->offset);outstr("(%edx)");nl();*/
   }
   else error("how to store?");
```

```
}
func rvalue(lval:*elval)
{
  if(!lval)
  {error("rvalue(0)");return;}
  if(lval->sort==L_NUM)
  loadnum(lval);
  else if(lval->sort==L_ID&&lval->idx)/*variable...*/
  getmem(lval);
  else if(lval->sort==L_POI)
  loadbyre(lval);
  else if(lval->sort==L_STR)
  {
    loadlita(lval);
  }
  else
  {
    fprintf(stderr,"lval->sort=%d\n",lval->sort);
    error("code generator error");
  }
}
func loadbyre(lval:*elval)
{
  if(typtab[lval->typ].size==BYTESIZE)/*ot("movsbl ");*/
  zlbrb(lval->offset);
  else if(typtab[lval->typ].size==WORDSIZE)/*ot("movl ");*/
  zlbrw(lval->offset);
  else if(typtab[lval->typ].sort==V_ARR)/*ot("leal ");*/
  zlbra(lval->offset);
  else error("error in loadbyreg()");
  /*outdec(lval->offset);
  outstr("(%eax), %eax");nl();*/
}
func getmem(lval:*elval)
{
  /*trc("getmem");*/
  if(lval->typ==T_CHAR)
  {
    /*ot("movsbl ");*/
    if(lval->idx->sort==S_VARG)
    {
      zldb(lval->name,lval->offset);
      /*outname(lval->name);
      if(lval->offset)
      {
        outasm("+");
        outdec(lval->offset);
      }
      outasm(", %eax");
```

```
    nl();*/
  }
  else if(lval->idx->sort==S_VARL)
  {
    zldlb(lval->idx->offset+lval->offset);
    /*outdec(lval->idx->offset+lval->offset);
    outasm("(%ebp), %eax");
    nl();*/
  }
  else error("error loading 'char' object");
}
else if(lval->typ==T_INT||lval->typ==T_INTP
    ||lval->typ==T_CHARP
    ||typtab[lval->typ].sort==V_PTR)
{
  /*ot("movl ");*/
  if(lval->idx->sort==S_VARG)
  {
    zldw(lval->name,lval->offset);
    /*outname(lval->name);
    if(lval->offset)
    {
      outasm("+");
      outdec(lval->offset);
    }
    outasm(", %eax");
    nl();*/
  }
  else if(lval->idx->sort==S_VARL)
  {
    zldlw(lval->idx->offset+lval->offset);
    /*outdec(lval->idx->offset+lval->offset);
    outasm("(%ebp), %eax");
    nl();*/
  }
  else error("getmem ?");
}
else if(typtab[lval->typ].sort==V_ARR)
{
  if(lval->idx->sort==S_VARG)
  {
    zlda(lval->name,lval->offset);
    /*ot("movl $");
    outname(lval->name);
    if(lval->offset)
    {
      outasm("+");
      outdec(lval->offset);
    }
```

```
      outasm(", %eax");nl();*/
    }
    else if(lval->idx->sort==S_VARL)
    {
      zlea(lval->idx->offset+lval->offset);
      /*ot("leal ");
      outdec(lval->idx->offset+lval->offset);
      outasm("(%ebp), %eax");nl();*/
    }
    else error("still: how to getmem?");
  }
  else error("again error");
}
func loadnum(lval:*elval)
{
  zldn(lval->val);
  /*ot("movl $");
  outdec(lval->val);
  outasm(", %eax");
  nl();*/
}
func expressi()
{
  var *enode:node;
  var elval:lval;
  node=bexptree();
  fprintf(stdout,"#:");pretree(node,stdout);fprintf(stdout,"\n");
  if(treetocode(node,&lval))rvalue(&lval);
  delenode(node);
}
func bexptree()
{
  var *enode:node;
  node=hcomma();
  return node;
}
func hcomma1()
{
  var *enode:node,newnode;
  node=hier1();
  blanks();
  if(ch()!=',')
  return node;
  while(1)
  {
    if(match(","))
    {
      newnode=getenode();
      newnode->l=node;
```

```
            newnode->r=hier1();
            newnode->third=0;
            newnode->op=OP_COMMA;
            node=newnode;
        }
        else return node;
    }
}
func hcomma()
{
    var *enode:node,newnode;
    node=hier1();
    blanks();
    if(ch()!=',')
    return node;
    while(1)
    {
        if(match(","))
        {
            newnode=getenode();
            newnode->l=node;
            newnode->r=hcomma();
            newnode->third=0;
            newnode->op=OP_COMMA;
            node=newnode;
        }
        else return node;
    }
}
func hier1()
{
    var *enode:node,newnode;
    node=hcond();
    if(match("="))
    {
        newnode=getenode();
        newnode->l=node;
        newnode->r=hier1();
        newnode->third=0;
        newnode->op=OP_ASSIGN;
        node=newnode;
    }
    return node;
}
func hcond()
{
    var *enode:node,newnode;
    node=hlor();
    if(match("?"))
```

```
  {
    newnode=getenode();
    newnode->l=node;
    newnode->r=hcond();
    blanks();
    if(!match(":"))
    error("':' expected i conditional");
    newnode->third=hcond();
    newnode->op=OP_COND;
    node=newnode;
  }
  return node;
}
func hlor()
{
  var *enode:node,newnode;
  node=hland();
  blanks();
  if(match("||"))
  {
    newnode=getenode();
    newnode->l=node;
    newnode->r=hlor();
    newnode->third=0;
    newnode->op=OP_LOR;
    node=newnode;
  }
  return node;
}
func h1lor()
{
  var *enode:node,newnode;
  node=hland();
  blanks();
  if(!streq(line+lptr,"||"))
  return node;
  while(match("||"))
  {
    newnode=getenode();
    newnode->l=node;
    newnode->r=hland();
    newnode->third=0;
    newnode->op=OP_LOR;
    node=newnode;
  }
  return node;
}
func hland()
{
```

```
  var *enode:node,newnode;
  node=hbor();
  blanks();
  if(match("&&"))
  {
    newnode=getenode();
    newnode->l=node;
    newnode->r=hland();
    newnode->third=0;
    newnode->op=OP_LAND;
    node=newnode;
  }
  return node;
}
func h1land()
{
  var *enode:node,newnode;
  node=hbor();
  blanks();
  if(!streq(line+lptr,"&&"))
  return node;
  while(match("&&"))
  {
    newnode=getenode();
    newnode->l=node;
    newnode->r=hbor();
    newnode->third=0;
    newnode->op=OP_LAND;
    node=newnode;
  }
  return node;
}
func hbor()
{
  var *enode:node,newnode;
  node=hbxor();
  if(ch()!='|'||streq(line+lptr,"||"))
  return node;
  while(1)
  {
    blanks();
    if(streq(line+lptr,"|")&&!streq(line+lptr,"||"))
    {
      gch();
      newnode=getenode();
      newnode->l=node;
      newnode->r=hbxor();
      newnode->third=0;
      newnode->op=OP_BOR;
```

```
        node=newnode;
      }
      else
      return node;
    }
}
func hbxor()
{
    var *enode:node,newnode;
    node=hband();
    blanks();
    if(ch()!='^'||streq(line+lptr,"^^"))
    return node;
    while(1)
    {
      blanks();
      if(streq(line+lptr,"^")&&!streq(line+lptr,"^^"))
      {
        gch();
        newnode=getenode();
        newnode->l=node;
        newnode->r=hband();
        newnode->third=0;
        newnode->op=OP_BXOR;
        node=newnode;
      }
      else return node;
    }
}
func hband()
{
    var *enode:node,newnode;
    node=hequal();
    blanks();
    if(ch()!='&'||streq(line+lptr,"&&"))
    return node;
    while(1)
    {
      blanks();
      if(ch()=='&'&&!streq(line+lptr,"&&"))
      {
        gch();
        newnode=getenode();
        newnode->l=node;
        newnode->r=hequal();
        newnode->third=0;
        newnode->op=OP_BAND;
        node=newnode;
      }
```

```
      else return node;
  }
}
func hequal()
{
  var *enode:node,newnode;
  node=hgt();
  blanks();
  if(!streq(line+lptr,"==")&&!streq(line+lptr,"!="))
  return node;
  while(1)
  {
    if(match("=="))
    {
      newnode=getenode();
      newnode->l=node;
      newnode->r=hgt();
      newnode->third=0;
      newnode->op=OP_EQ;
      node=newnode;
    }
    else if(match("!="))
    {
      newnode=getenode();
      newnode->l=node;
      newnode->r=hgt();
      newnode->third=0;
      newnode->op=OP_NEQ;
      node=newnode;
    }
    else return node;
  }
}
func hgt()
{
  var *enode:node,newnode;
  node=hshift();
  blanks();
  if(!streq(line+lptr,"<")
   &&!streq(line+lptr,">")
   &&!streq(line+lptr,"<=")
   &&!streq(line+lptr,">=")
   ||streq(line+lptr,"<<")
   ||streq(line+lptr,">>"))
  return node;
  while(1)
  {
    if(match("<="))
    {
```

```
        newnode=getenode();
        newnode->l=node;
        newnode->r=hshift();
        newnode->third=0;
        newnode->op=OP_LE;
        node=newnode;
      }
      else if(match(">="))
      {
        newnode=getenode();
        newnode->l=node;
        newnode->r=hshift();
        newnode->third=0;
        newnode->op=OP_GE;
        node=newnode;
      }
      else if(streq(line+lptr,"<")&&!streq(line+lptr,"<<"))
      {
        gch();
        newnode=getenode();
        newnode->l=node;
        newnode->r=hshift();
        newnode->third=0;
        newnode->op=OP_LT;
        node=newnode;
      }
      else if(streq(line+lptr,">")&&!streq(line+lptr,">>"))
      {
        gch();
        newnode=getenode();
        newnode->l=node;
        newnode->r=hshift();
        newnode->third=0;
        newnode->op=OP_GT;
        node=newnode;
      }
      else return node;
   }
}
func hshift()
{
  var *enode:node,newnode;
  node=hplusminus();
  blanks();
  if(!streq(line+lptr,"<<")&&!streq(line+lptr,">>"))
  return node;
  while(1)
  {
    if(match("<<"))
```

```
      {
        newnode=getenode();
        newnode->l=node;
        newnode->r=hplusminus();
        newnode->third=0;
        newnode->op=OP_SHL;
        node=newnode;
      }
      else if(match(">>"))
      {
        newnode=getenode();
        newnode->l=node;
        newnode->r=hplusminus();
        newnode->third=0;
        newnode->op=OP_SHR;
        node=newnode;
      }
      else return node;
  }
}
func hplusminus()
{
  var *enode:node,newnode;
  node=hmuldiv();
  blanks();
  if((ch()!='+')&&(ch()!='-'))
  return node;
  while(1)
  {
    if(match("+"))
    {
      newnode=getenode();
      newnode->l=node;
      newnode->r=hmuldiv();
      newnode->third=0;
      newnode->op=OP_PLUS;
      node=newnode;
    }
    else if(match("-"))
    {
      newnode=getenode();
      newnode->l=node;
      newnode->r=hmuldiv();
      newnode->third=0;
      newnode->op=OP_MINUS;
      node=newnode;
    }
    else
    return node;
```

```
    }
}
func hmuldiv()
{
   var *enode:node,newnode;
   node=hunary();
   blanks();
   if((ch()!='*')&&(ch()!='/')&&(ch()!='%'))
   return node;
   while(1)
   {
     if(match("*"))
      {
        newnode=getenode();
        newnode->l=node;
        newnode->r=hunary();
        newnode->third=0;
        newnode->op=OP_MUL;
        node=newnode;
      }
     else if(match("/"))
      {
        newnode=getenode();
        newnode->l=node;
        newnode->r=hunary();
        newnode->third=0;
        newnode->op=OP_DIV;
        node=newnode;
      }
     else if(match("%"))
      {
        newnode=getenode();
        newnode->l=node;
        newnode->r=hunary();
        newnode->third=0;
        newnode->op=OP_REM;
        node=newnode;
      }
     else
     return node;
   }
}
func hunary()
{
   var *enode:node,newnode;
   if(match("++"))
   {
     node=hunary();
     newnode=getenode();
```

```
          newnode->l=0;
          newnode->r=node;
          newnode->third=0;
          newnode->op=OP_1PP;
          node=newnode;
          return node;
        }
        else if(match("--"))
        {
          node=hunary();
          newnode=getenode();
          newnode->l=0;
          newnode->r=node;
          newnode->third=0;
          newnode->op=OP_1MM;
          node=newnode;
          return node;
        }
        else if(match("+"))
        {
          node=hunary();
          return node;
        }
        else if(ch()=='-'&&!numeric(line[lptr+1])/*match("-")*/)
        {
          gch();
          node=hunary();
          newnode=getenode();
          newnode->l=0;
          newnode->r=node;
          newnode->third=0;
          newnode->op=OP_UMINUS;
          node=newnode;
          return node;
        }
        else if(match("*"))
        {
          node=hunary();
          newnode=getenode();
          newnode->l=0;
          newnode->r=node;
          newnode->third=0;
          newnode->op=OP_STAR;
          node=newnode;
          return node;
        }
        else if(match("!"))
        {
          node=hunary();
```

```
    newnode=getenode();
    newnode->l=0;
    newnode->r=node;
    newnode->third=0;
    newnode->op=OP_LNOT;
    node=newnode;
    return node;
}
else if(match("~"))
{
    node=hunary();
    newnode=getenode();
    newnode->l=0;
    newnode->r=node;
    newnode->third=0;
    newnode->op=OP_BNOT;
    node=newnode;
    return node;
}
else if(match("&"))
{
    node=hunary();
    newnode=getenode();
    newnode->l=0;
    newnode->r=node;
    newnode->third=0;
    newnode->op=OP_ADDR;
    node=newnode;
    return node;
}
else
{
    node=hsubscr();
    if(match("++"))
    {
        newnode=getenode();
        newnode->l=0;
        newnode->r=node;
        newnode->third=0;
        newnode->op=OP_2PP;
        node=newnode;
    }
    else if(match("--"))
    {
        newnode=getenode();
        newnode->l=0;
        newnode->r=node;
        newnode->third=0;
        newnode->op=OP_2MM;
```

```
        node=newnode;
      }
      return node;
  }
}
func hsubscr()
{
  var *enode:node,newnode;
  node=primary();
  blanks();
  if((ch()==']')||
   streq(line+lptr,tlarg)
   ||(ch()=='.')
   ||streq(line+lptr,"->"))
  while (1)
    {
    if(match("["))
      {/* a[i] --> *((a)+(i)) */
       newnode=getenode();
       newnode->l=node;
       newnode->r=hcomma();
       needbrac("]");
       newnode->third=0;
       newnode->op=OP_PLUS;
       node=getenode();
       node->l=0;
       node->r=newnode;
       node->third=0;
       node->op=OP_STAR;
       }
    else if(match(tlarg/*"("*/))
      {
       newnode=getenode();
       newnode->l=node;
       newnode->r=0;
       newnode->third=0;
       newnode->op=OP_FUNC;
       node=newnode;
       blanks();
       while(!streq(line+lptr,trarg))
         {
         if(endst())break;
         newnode=getenode();
         newnode->l=hier1();
         newnode->r=node->r;
         newnode->third=0;
         newnode->op=OP_LIST;
         node->r=newnode;
         if(!match(","))break;
```

```
          }
        needbrac(trarg/*")"*/);
        }
      else if(match("->"))
        {/* p->a (*p).a */
        var [NAMESIZE]char:nm;
        if(!symname(nm))
          error("field name expected");
        newnode=getenode();
        newnode->l=0;
        newnode->r=node;
        newnode->third=0;
        newnode->op=OP_STAR;
        node=getenode();
        node->l=newnode;
        node->r=0;
        node->third=0;
        node->op=OP_DOT;
        strcp(node->name,nm);
        }
      else if(match("."))
        {
        var [NAMESIZE]char:nm;
        if(!symname(nm))
          error("field name expected");
        newnode=getenode();
        newnode->l=node;
        newnode->r=0;
        newnode->third=0;
        newnode->op=OP_DOT;
        strcp(newnode->name,nm);
        node=newnode;
        }
      else
        return node;
      }
  return node;
}
func getenode()
{
  return chkmem(calloc(1,sizeof(enode)));
}
func delenode(node:*enode)
{
  if(!node)return;
  delenode(node->l);
  delenode(node->r);
  delenode(node->third);
  free(node);
```

```
}
func primary()
{
  var *enode:node;
  var [1]int: num;
  var [NAMESIZE]char: nm;
  if(match(tlarg))
  {
    node=bexptree();
    needbrac(trarg);
    return node;
  }
  if(amatch("sizeof",6))
  {
    var int:t,s;
    needbrac(tlarg);
    t=gettypen();
    s=gettsize(t);
    needbrac(trarg);
    node=getenode();
    node->l=node->r=node->third=0;
    node->op=OP_LEAF;
    node->leaf.vid=L_NUM;
    node->leaf.val=s;
    return node;
  }
  if(number(num))
  {
    node=getenode();
    node->l=node->r=node->third=0;
    node->op=OP_LEAF;
    node->leaf.vid=L_NUM;
    node->leaf.val=num[0];
    return node;
  }
  if(pstr(num))
  {
    node=getenode();
    node->l=node->r=node->third=0;
    node->op=OP_LEAF;
    node->leaf.vid=L_NUM;
    node->leaf.val=num[0];
    return node;
  }
  if(symname(nm))
  {
    var *ssym:k;
    node=getenode();
    node->l=node->r=node->third=0;
```

```
    node->op=OP_LEAF;
    node->leaf.vid=L_ID;
    strcp(node->name,nm);
    if(methodcls&&methodidx&&(k=findfiel(nm,methodcls)))
    {
       var *enode:ths;
       var *ssym:t;
       ths=getenode();
       ths->l=ths->r=ths->third=0;
       ths->op=OP_LEAF;
       ths->leaf.vid=L_ID;
       if(!(t=findloc("this")))
       error("internal error: missing 'this'");
       ths->leaf.idx=t;
       strcp(ths->name,"this");
       node->r=ths;
       node->op=OP_STAR;
       var *enode:newnode;
       newnode=getenode();
       newnode->l=node;
       newnode->r=0;
       newnode->third=0;
       newnode->op=OP_DOT;
       strcp(newnode->name,nm);
       node=newnode;
    }
    else if(k=findloc(nm))
    {
       node->leaf.idx=k;
       cnmlst->addm(nm);
    }
    else if(k=findglb(nm))
    {
       node->leaf.idx=k;
       cnmlst->addm(nm);
    }
    else
    {/* undeclared function */
       k=addloc(nm,S_VARL,1,0,T_INT);
       k->sort=S_FUNC;
       k->dfd=0;
       node->leaf.idx=k;
       cnmlst->addm(nm);
    }
    return node;
}
if(qstr(num))
{
    node=getenode();
```

```
      node->l=node->r=node->third=0;
      node->op=OP_LEAF;
      node->leaf.vid=L_STR;
      node->leaf.val=num[0];
      return node;
   }
   error("wrong expression");
   return 0;
}
func pstr(val:*int)
{
   var int:k;
   var char:c;
   k=0;
   if(!match("'"))return 0;
   while((c=gch())!=39)
   {
      if(c!=92)
      k=((k&255)<<8)+(c&255);
      else
      {
         c=gch();
         if(!c)break;
         if(c=='n')c=10;
         else if(c=='t')c=9;
         else if(c=='b')c=8;
         else if(c=='f')c=12;
         k=((k&255)<<8)+(c&255);
      }
   }
   val[0]=k;
   /*fprintf(stderr,"the character:%d\n",k);*/
   return 1;
}
func qstr(val:*int)
{
   var int:k;
   var char:c;
   k=0;
   if(!match(quote))return 0;
   val[0]=stptr;
   while(ch()!='"')
   {
      if(!ch())break;
      if(stptr>=STMAX)
      {
         error("string space exhausted");
         while(!match(quote))if(!gch())break;
         return 1;
```

```
        }
      c=gch();
      if(c!=92)litq[stptr++]=c;
      else
        {
          c=gch();
          if(!c)break;
          if(c=='n')c=10;
          else if(c=='t')c=9;
          else if(c=='b')c=8;
          else if(c=='f')c=12;
          litq[stptr++]=c;
        }
    }
  gch();
  litq[stptr++]=0;
  return 1;
}
func number(val:*int)
{
  var int:k;
  var int:d;
  var int:minus;
  var char:c;
  k=minus=1;
  if(match("0x")){
  k=0;
  while(numeric(ch())||((ch()>='a')&&(ch()<='f'))){
    c=inbyte();
    if(numeric(c)){
    k=(k<<4)+(c-'0');
    }
    else
    {
      k=(k<<4)+(c-'a'+10);
    }
  }
  val[0]=k;
  return 1;
  }
  while(k){
  k=0;
  if(match("+"))k=1;
  if(match("-")){
    minus=-1;
    k=1;
  }
  }
  if(!numeric(ch()))return 0;
```

```
    while(numeric(ch())){
    c=inbyte();
    k=k*10+(c-'0');
    }
    if(minus<0)k=(-k);
    val[0]=k;
    return 1;
}
func needbrac(p:*char)
{
    if(!match(p)){
    error("missing bracket");
    comment();
    outstr(p);
    nl();
    }
}
func needlval()
{
    error("must be lvalue");
}
func ns()
{
    if(!match(";"))error("missing ';'");
}
func issigned(t:int)
{
    return (t==T_INT)||(t==T_CHAR);
}
func gettsize(t:int)
{
    if(t==T_INT)return WORDSIZE;
    else if(t==T_CHAR)return BYTESIZE;
    else if(t==T_INTP)return WORDSIZE;
    else if(t==T_CHARP)return WORDSIZE;
    else if((t>=F_TYPE)&&(t<typptr))return typtab[t].size;
    else
    {
      error("strange type");
      return 0;
    }
}
/*returns type 'function of a type' */
func getfnctype(t:int)
{
    if(t>=typptr)
    {
      error("getfnctype:unknown type");
      return T_INTP;
```

```
    }
    var int:k;
    for(k=1;k<typptr;k++)
    if((typtab[k].sort==V_FNC)&&(typtab[k].type==t))return k;
    k=typget();
    typtab[k].name[0]=0;
    typtab[k].sort=V_FNC;
    typtab[k].type=t;
    typtab[k].size=0;/*maybe...*/
    return k;
}
func typget()
{
    if(typptr>=NUMTYP)
    {
        numtyp=numtyp+NUMTYP;
        chkmem(typtab=realloc(typtab,sizeof(styp)*numtyp));
    }
    return typptr++;
}
/*returns pointer to a type */
func getptrty(t:int)
{
    var int:k;
    if(t>=typptr){
    error("unknown type");
    return T_INTP;
    }
    k=1;
    while(k<typptr){
    if((typtab[k].sort==V_PTR)&&(typtab[k].type==t))return k;
    k++;
    }
    if(typptr>=numtyp){
    /*fprintf(stderr,"reallocating types\n");*/
    numtyp=numtyp+NUMTYP;
    chkmem(typtab=realloc(typtab,sizeof(styp)*numtyp));
    }
    typtab[typptr].name[0]=0;
    typtab[typptr].sort=V_PTR;
    typtab[typptr].type=t;
    typtab[typptr].size=WORDSIZE;
    return typptr++;
}
func cbtype()
{
    var [NAMESIZE]char:sname;
    blanks();
    /*fprintf(stderr,"cbtype()\n");
```

```
      fprintf(stderr,"line=%s\n",line);*/
      if(ch()=='*'||ch()=='[')
      return 1;
      if(!fsymname(sname))
      return 0;
      /*fprintf(stderr,"got a name:%s\n,line=%s\n",sname,line);*/
      if(!findtyp(sname))
      return 0;
      return 1;
}
func gettypen()
{
   var [NAMESIZE]char:tname;
   var int:k;
   var int:l;
   var int:t;
   var int:c;
   var [1]int:dim;
   trc("gettypename");
   if(amatch("char",4))return T_CHAR;
   if(amatch("int",3))return T_INT;
   if(alpha(ch())){
   k=l=0;
   while(an(c=line[lptr+k])){
      if(k<NAMEMAX)tname[l++]=c;
      k++;
   }
   comment();
   outstr("l=");
   outdec(l);
   nl();
   tname[l]=0;
   comment();
   outstr("typename:");
   outstr(tname);
   nl();
   if(t=findtyp(tname)){
      lptr=lptr+k;
      comment();
      outstr("type handle:");
      outdec(t);
      nl();
      trc(line+lptr);
      return t;
   }
   else
      {
      error("unknown type");
      return T_INT;
```

```
    }
  }
if(match("*")){
if(!(t=gettypen()))return 0;
k=1;
while(k<typptr){
    if((typtab[k].sort==V_PTR)&&(typtab[k].type==t))return k;
    k++;
  }
if(typptr>=numtyp){
    /*fprintf(stderr,"reallocating types\n");*/
    numtyp=numtyp+NUMTYP;
    chkmem(typtab=realloc(typtab,sizeof(styp)*numtyp));
    }
typtab[typptr].name[0]=0;
typtab[typptr].sort=V_PTR;
typtab[typptr].type=t;
typtab[typptr].size=WORDSIZE;
return typptr++;
}
if(match("[")){
if(!number(dim)){
    error("dimension expected");
    dim[0]=1;
  }
needbrac("]");
if(!(t=gettypen()))return 0;
k=1;
while(k<typptr){
    if((typtab[k].sort==V_ARR)&&(typtab[k].type==t)
     &&(typtab[k].dim==dim[0]))return k;
    k++;
  }
if(typptr>=numtyp){
    /*fprintf(stderr,"reallocating types\n");*/
    numtyp=numtyp+NUMTYP;
    chkmem(typtab=realloc(typtab,sizeof(styp)*numtyp));
  }
typtab[typptr].name[0]=0;
typtab[typptr].sort=V_ARR;
typtab[typptr].type=t;
typtab[typptr].size=gettsize(t)*dim[0];
typtab[typptr].dim=dim[0];
return typptr++;
  }
error("type expected");
return T_INT;
}
func outname(n:*char)
```

```
{
  outasm(n);
}
func inbyte()
{
  while(!ch()){
  if(iseof)return 0;
  insline();
  preproce();
  }
  return gch();
}
func junk()
{
  if(an(inbyte()))while(an(ch()))gch();
  else
  while(!an(ch())){
    if(!ch())break;
    gch();
  }
  blanks();
}
func strcp(d:*char,s:*char)
{
  while(*d++=*s++);
}
func findsym(sname:*char,syms:*ssym,ptr:int)
{
  var int:k;
  /*fprintf(stderr,"sname=%s,ptr=%d\n",sname,ptr);*/
  for(k=0;k<ptr;k++)
  {
    /*fprintf(stderr,"1");*/
    if(strid(sname,syms[k].name))
    {return syms+k;fprintf(stderr,"2\n");}
    /*fprintf(stderr,"2");*/
  }
  return 0;
}
func outdec(n:int)
{
  if(n<0){
  outbyte('-');
  n=-n;
  }
  outint(n);
}
func outint(n:int)
{
```

```
    var int:q;
    q=n/10;
    if(q)outint(q);
    outbyte('0'+n%10);
}
func error(p:*char)
{
    var int:k;
    comment();
    outstr("Error:");
    outstr(p);
    fprintf(stderr,"---->Error:%s\n",p);
    fprintf(stderr,"%s\n",line);
    for(k=lptr;k--;)fprintf(stderr," ");
    fprintf(stderr,"^\n");
    nl();
    ++errcnt;
}
func reset()
{
    lptr=0;
    line[0]=0;
}
func ol(p:*char)
{
    ot(p);
    nl();
}
func dumplits()
{
    var int:j;
    var int:k;
    if(!stptr)return ;
    ot(".section");
    ot(".rodata");
    nl();
    printlab(stlab);
    col();
    nl();
    k=0;
    while(k<stptr){
    defbyte();
    j=10;
    while(j--){
        outdec(litq[k++]);
        if(!j|(k>=stptr)){
        nl();
        break;
        }
```

```
      outbyte(',');
  }
  }
}
func defbyte()
{
  ot(".byte ");
}
func defstora()
{
  ot(".comm ");
}
func printmap()
{
  var of:*int;
  if(!(of=fopen(mapname,"w")))
  {
    fprintf(stderr,"can't open map file %s\n",mapname);
    return;
  }
  var int:s;
  var *ssymlist:lst;
  for(lst=glbsymtab.lst;lst;lst=lst->next)
  {
    fprintf(of,"%-10s\t%d\t%d\n",lst->sym.name,
        lst->sym.line,
        lst->sym.sort);
  }
  fclose(of);
}
func dumpglbs()
{
  var int:s;
  var *ssymlist:lst;
  for(lst=glbsymtab.lst;lst;lst=lst->next)
  {
    if(lst->sym.sort==S_VARG)if(lst->sym.dfd)
    {
      defstora();
      outname(lst->sym.name);
      comma();
      s=gettsize(lst->sym.type);
      outdec(s);
      if(lst->sym.type==T_CHAR)outasm(",1");
      else
      outasm(",4");
      nl();
    }
  }
```

```
}
func trailer()
{
  nl();
  comment();
  outstr("<<End of compilation>>");
  nl();
  ot(".ident");
  tab();
  outstr(quote);
  outstr("T Cmplr");
  outstr(quote);
  nl();
  /*tstcg();*/
}
func nl()
{
  outbyte(10);
}
func tab()
{
  outbyte(9);
}
func col()
{
  outbyte(':');
}
func comma()
{
  outbyte(',');
}
func comment()
{
  outbyte('#');
  outbyte(':');
}
func ot(p:*char)
{
  tab();
  outasm(p);
}
func outbyte(c:char)
{
  if(!tolitstk)return outbyte1(c);
  return putlitst(c);
}
func outbyte1(c:char)
{
  if(!c)return 0;
```

```
    putchar(c);
    return c;
}
func outasm(p:*char)
{
    while(outbyte(*p++));
}
func outstr(p:*char)
{
    while(*p)outbyte(*p++);
}
func symname(sname:*char)
{
    var int:k;
    var int:l;
    var char:c;
    blanks();
    if(!alpha(ch()))return 0;
    k=l=0;
    while(an(ch()))
    {
        if(k<NAMEMAX)
        {
            sname[l++]=gch();
            k++;
        }
        else
        {
            k++;
            gch();
        }
    }
    sname[l]=0;
    return 1;
}
func fsymname(sname:*char)
{
    var int:k;
    var int:l;
    var char:c;
    var int:qptr;
    blanks();
    if(!alpha(ch()))return 0;
    k=l=0;
    qptr=lptr;
    /*fprintf(stderr,"%s\n",line);*/
    while(an(line[qptr]))
    {
        /*fprintf(stderr,"%s\n",line);*/
```

```
    if(k<NAMEMAX)
    {
      sname[l++]=line[qptr++];
      k++;
    }
    else
    {
      k++;
      qptr++;
    }
  }
  sname[l]=0;
  return 1;
}
func trc(p:*char)
{
  comment();
  outstr(p);
  nl();
}
```

## 2.2  codegen.e

```
/*                    -*- C -*-                              */
/*              code generation by E.V., (C) 2003           */
/*
  This program is free software; you can redistribute it and/or modify
  it under the terms of the GNU General Public License as published by
  the Free Software Foundation; either version 2 of the License, or
  (at your option) any later version.

  This program is distributed in the hope that it will be useful,
  but WITHOUT ANY WARRANTY; without even the implied warranty of
  MERCHANTABILITY or FITNESS FOR A PARTICULAR PURPOSE.  See the
  GNU General Public License for more details.

  You should have received a copy of the GNU General Public License
  along with this program; if not, write to the Free Software
  Foundation, Inc., 59 Temple Place, Suite 330, Boston, MA  02111-1307  USA
*/
#include "codegen.he"
var extern *int:stderr;
var extern int:stlab;

func cg_init(this:*scodegen)
{
  this->ncodeitems=NCODEITEMS;
  chkmem(this->codes=calloc(this->ncodeitems,sizeof(scode)));
  this->codeptr=0;
}
```

```
func cg_done(this:*scodegen)
{
  if(this->codes)
  {
    var i:int;
    for(i=0;i<this->codeptr;i++)
    cd_done(this->codes+i);
    free(this->codes);
  }
  this->codes=0;
  this->codeptr=0;
  this->ncodeitems=0;
}
func cg_getitem(this :*scodegen)
{
  if(this->codeptr>=this->ncodeitems)
  {
    this->ncodeitems=this->ncodeitems+NCODEITEMS;
    /*fprintf(stderr,"ncodeitems=%d, reallocating\n",
    this->ncodeitems);*/
    chkmem(this->codes=realloc(this->codes,
                  this->ncodeitems*sizeof(scode)));
  }
  cd_init(this->codes+this->codeptr);
  return this->codes+this->codeptr++;
}
func cg_print(*scodegen this)
{
  var int:i;
  for(i=0;i<this->codeptr;i++)
  cd_write(this->codes+i);
}
func cg_transfer(*scodegen this,*scodegen dest)
{
  /*fprintf(stderr,"transfer(),dest=%d\n",dest);*/
  var int:i;
  for(i=0;i<this->codeptr;++i)
  cg_insert(dest,this->codes+i);
  /*fprintf(stderr,"!transfer()\n");*/
}
func cg_insert(*scodegen this,*scode s)
{
  /*fprintf(stderr,"insert()\n");*/
  var *scode:d;
  d=cg_getitem(this);
  d->code=s->code;
  d->arg=s->arg;
  d->reg=s->reg;
  if(s->str)
```

```
    d->str=strdyn(s->str);
    else
    d->str=0;
    /*fprintf(stderr,"!insert()\n");*/
}
func cd_write(*scode:this)
{
    if(this->code==CD_ZCALL)
    {
        ot("call ");
        outname(this->str);
        nl();
    }
    else if(this->code==CD_LAB)
    {
        printlab(this->arg);
        col();
        nl();
    }
    else if(this->code==CD_JUMP)
    {
        ot("jmp ");
        printlab(this->arg);
        nl();
    }
    else if(this->code==CD_TESTJUMP)
    {
        ol("testl %eax, %eax");
        ot("je");
        tab();
        printlab(this->arg);
        nl();
    }
    else if(this->code==CD_TESTNEJUMP)
    {
        ol("testl %eax, %eax");
        ot("jne");
        tab();
        printlab(this->arg);
        nl();
    }
    else if(this->code==CD_NEG)
    {
        ol("negl %eax");
    }
    else if(this->code==CD_LNOT)
    {
        ol("testl %eax,%eax");
        ol("sete %al");
```

```
  ol("movzbl %al, %eax");
}
else if(this->code==CD_BNOT)
{
  ol("notl %eax");
}
else if(this->code==CD_EQ)
{
  ot("cmpl");
  ot("%eax, %edx");
  nl();
  ot("sete");
  ot("%al");
  nl();
  ot("movzbl");
  ot("%al, %eax");
  nl();
}
else if(this->code==CD_NEQ)
{
  ot("cmpl");
  ot("%eax, %edx");
  nl();
  ot("setne");
  ot("%al");
  nl();
  ot("movzbl");
  ot("%al, %eax");
  nl();
}
else if(this->code==CD_ZGE)
{
  ot("cmpl");
  ot("%eax, %edx");
  nl();
  ot("setge");
  ot("%al");
  nl();
  ot("movzbl");
  ot("%al, %eax");
  nl();
}
else if(this->code==CD_UGE)
{
  ot("cmpl");
  ot("%eax, %edx");
  nl();
  ot("setae");
  ot("%al");
```

```
  nl();
  ot("movzbl");
  ot("%al, %eax");
  nl();
}
else if(this->code==CD_ZLE)
{
  ot("cmpl");
  ot("%eax, %edx");
  nl();
  ot("setle");
  ot("%al");
  nl();
  ot("movzbl");
  ot("%al, %eax");
  nl();
}
else if(this->code==CD_ULE)
{
  ot("cmpl");
  ot("%eax, %edx");
  nl();
  ot("setbe");
  ot("%al");
  nl();
  ot("movzbl");
  ot("%al, %eax");
  nl();
}
else if(this->code==CD_ZLT)
{
  ot("cmpl");
  ot("%eax, %edx");
  nl();
  ot("setl");
  ot("%al");
  nl();
  ot("movzbl");
  ot("%al, %eax");
  nl();
}
else if(this->code==CD_ULT)
{
  ot("cmpl");
  ot("%eax, %edx");
  nl();
  ot("setb");
  ot("%al");
  nl();
```

```
    ot("movzbl");
    ot("%al, %eax");
    nl();
}
else if(this->code==CD_ZGT)
{
    ot("cmpl");
    ot("%eax, %edx");
    nl();
    ot("setg");
    ot("%al");
    nl();
    ot("movzbl");
    ot("%al, %eax");
    nl();
}
else if(this->code==CD_UGT)
{
    ot("cmpl");
    ot("%eax, %edx");
    nl();
    ot("seta");
    ot("%al");
    nl();
    ot("movzbl");
    ot("%al, %eax");
    nl();
}
else if(this->code==CD_BOR2REGS)
{
    ol("orl %edx, %eax");
}
else if(this->code==CD_BXOR2REGS)
{
    ol("xorl %edx, %eax");
}
else if(this->code==CD_BAND2REGS)
{
    ol("andl %edx, %eax");
}
else if(this->code==CD_ADD2REGS)
{
    ol("addl %edx, %eax");
}
else if(this->code==CD_SUB2REGS)
{
    ol("subl %eax, %edx");
    ol("movl %edx, %eax");
}
```

```
else if(this->code==CD_MUL2REGS)
{
  ol("imull %edx");
}
else if(this->code==CD_DIV2REGS)
{
  ol("xchgl %eax, %edx");
  ol("movl %edx, %ecx");
  ol("cltd");
  ol("idivl %ecx");
}
else if(this->code==CD_MOD2REGS)
{
  ol("xchgl %eax, %edx");
  ol("movl %edx, %ecx");
  ol("cltd");
  ol("idivl %ecx");
  ol("movl %edx, %eax");
}
else if(this->code==CD_STKENTER)
{
  ol("pushl %ebp");
  ol("movl %esp, %ebp");
}
else if(this->code==CD_STKLEAVE)
{
  ol("movl %ebp, %esp");
  ol("popl %ebp");
}
else if(this->code==CD_INCREG)
{
  if(this->arg>0)
  {
    if(this->arg<3)
    while(this->arg--)
      ol("incl %eax");
    else
    {
      ot("addl $");
      outdec(this->arg);
      outstr(", %eax");
      nl();
    }
  }
}
else if(this->code==CD_DECREG)
{
  if(this->arg>0)
  {
```

```
      if(this->arg<3)
      while(this->arg--)
        ol("decl %eax");
      else
      {
        ot("subl $");
        outdec(this->arg);
        outstr(", %eax");
        nl();
      }
    }
  }
  else if(this->code==CD_MODSTK)
  {
    if(this->arg>0)
    {
      ot("addl $");
      outdec(this->arg);
      outasm(", %esp");
      nl();
    }
    else if(this->arg<0)
    {
      ot("subl $");
      outdec(-this->arg);
      outasm(", %esp");
      nl();
    }
  }
  else if(this->code==CD_SHL)
  {
    ol("movl %eax, %ecx");
    ol("movl %edx, %eax");
    ol("sall %cl, %eax");
  }
  else if(this->code==CD_ASR)
  {
    ol("movl %eax, %ecx");
    ol("movl %edx, %eax");
    ol("sarl %cl, %eax");
  }
  else if(this->code==CD_SHR)
  {
    ol("movl %eax, %ecx");
    ol("movl %edx, %eax");
    ol("shrl %cl, %eax");
  }
  else if(this->code==CD_MULREG)
  {
```

```
   xmulreg(this->arg,this->reg[regnames]);
}
else if(this->code==CD_DIVCONST)
{
   xdivconst(this->arg);
}
else if(this->code==CD_PUSH)
{
   ol("pushl %eax");
}
else if(this->code==CD_POP)
{
   ol("popl %edx");
}
else if(this->code==CD_RET)
{
   ol("ret");
}
else if(this->code==CD_LDLIT)
{
   ot("movl $");
   printlab(stlab);
   outstr("+");
   outdec(this->arg);
   outstr(", %eax");
   nl();
}
else if(this->code==CD_LDN)
{
   if(this->arg==0)
   ol("xorl %eax, %eax");
   else
   {
     ot("movl $");
     outdec(this->arg);
     outstr(", %eax");
     nl();
   }
}
else if(this->code==CD_LDA)
{
   ot("movl $");
   outname(this->str);
   if(this->arg)
   {outasm("+");outdec(this->arg);}
   outasm(", %eax");
   nl();
}
else if(this->code==CD_LEA)
```

```
{
  ot("leal ");
  outdec(this->arg);
  outasm("(%ebp), %eax");
  nl();
}
else if(this->code==CD_STOW)
{
  ot("movl %eax, ");
  outname(this->str);
  if(this->arg)
  {outstr("+");outdec(this->arg);}
  nl();
}
else if(this->code==CD_STOB)
{
  ot("movb %al, ");
  outname(this->str);
  if(this->arg)
  {outstr("+");outdec(this->arg);}
  nl();
}
else if(this->code==CD_STOB2)
{
  ot("movb %al, ");
  outdec(this->arg);
  outstr("(%edx)");nl();
}
else if(this->code==CD_STOW2)
{
  ot("movl %eax, ");
  outdec(this->arg);
  outstr("(%edx)");nl();
}
else if(this->code==CD_STLW)
{
  ot("movl %eax, ");
  outdec(this->arg);outasm("(%ebp)");nl();
}
else if(this->code==CD_STLB)
{
  ot("movb %al, ");
  outdec(this->arg);outasm("(%ebp)");nl();
}
else if(this->code==CD_LBRB)
{
  ot("movsbl ");
  outdec(this->arg);
  outstr("(%eax), %eax");nl();
```

```
}
else if(this->code==CD_LBRW)
{
  ot("movl ");
  outdec(this->arg);
  outstr("(%eax), %eax");nl();
}
else if(this->code==CD_LBRA)
{
  ot("leal ");
  outdec(this->arg);
  outstr("(%eax), %eax");nl();
}
else if(this->code==CD_LDW)
{
  ot("movl ");
  outname(this->str);
  if(this->arg)
  {
    outasm("+");
    outdec(this->arg);
  }
  outasm(", %eax");nl();
}
else if(this->code==CD_LDB)
{
  ot("movsbl ");
  outname(this->str);
  if(this->arg)
  {
    outasm("+");
    outdec(this->arg);
  }
  outasm(", %eax");nl();
}
else if(this->code==CD_LDLW)
{
  ot("movl ");
  outdec(this->arg);
  outasm("(%ebp), %eax");
  nl();
}
else if(this->code==CD_LDLB)
{
  ot("movsbl ");
  outdec(this->arg);
  outasm("(%ebp), %eax");
  nl();
}
```

```
  else if(this->code==CD_IGNORE)
  ;
  else
  {
    fprintf(stderr,"%d ",this->code);
    error("unknown opcode");
  }
}
func xdivconst(k:int)
{
  var int:l;
  if(k==1)return ;
  if(!k){
  error("division by zero");
  return ;
  }
  l=1;
  while(l<15)if(k==(1<<l)){
  ot("sarl $");
  outdec(l);
  outstr(", %eax");
  nl();
  return ;
  }
  else
  l++;
  ol("cltd");
  ot("divl $");
  outdec(k);
  nl();
}
func xmulreg(k:int,s:*char)
{
  var int:l;
  if(k==1)return ;
  else
  if(k==0){
    ot("xorl ");
    outstr(s);
    outstr(", ");
    outstr(s);
    nl();
  }
  else
    {
    l=1;
    while(l<15)if(k==(1<<l)){
      ot("sall $");
      outdec(l);
```

```
        outstr(", ");
        outstr(s);
        nl();
        return ;
      }
      else
        l++;
      }
  ot("imull $");
  outdec(k);
  outstr(", ");
  outstr(s);
  nl();
}
func cd_init(*scode:this)
{
  this->code=0;
  this->arg=0;
  this->reg=RG_A;
  this->str=0;
}
func cd_done(this:*scode)
{
  if(this->str)
  free(this->str);
  this->code=0;
  this->arg=0;
  this->reg=RG_A;
  this->str=0;
}
func strdyn(*char:s)
{
  var int :l;
  var *char:res;
  l=strlen(s);
  chkmem(res=calloc(l+1,sizeof(char)));
  strcp(res,s);
  return res;
}
func mulreg(k:int,int s)
{
  var *scode:cd;
  cd=cg_getitem(ccg);
  cd->code=CD_MULREG;
  cd->arg=k;
  cd->reg=s;
}
func increg(k:int)
{
```

```
   var *scode:cd;
   cd=cg_getitem(ccg);
   cd->code=CD_INCREG;
   cd->arg=k;
}
func decreg(k:int)
{
   var *scode:cd;
   cd=cg_getitem(ccg);
   cd->code=CD_DECREG;
   cd->arg=k;
}
func mult()
{
   var *scode:cd;
   cd=cg_getitem(ccg);
   cd->code=CD_MUL2REGS;
}
func div()
{
   var *scode:cd;
   cd=cg_getitem(ccg);
   cd->code=CD_DIV2REGS;
}
func zmod()
{
   var *scode:cd;
   cd=cg_getitem(ccg);
   cd->code=CD_MOD2REGS;
}
func zpop()
{
   var *scode:cd;
   cd=cg_getitem(ccg);
   cd->code=CD_POP;
   Zsp=Zsp+4;
}
func zpush()
{
   var *scode:cd;
   cd=cg_getitem(ccg);
   cd->code=CD_PUSH;
   Zsp=Zsp-4;
}
func zret()
{
   var *scode:cd;
   cd=cg_getitem(ccg);
   cd->code=CD_RET;
```

```
}
func divconst(int k)
{
  var *scode:cd;
  cd=cg_getitem(ccg);
  cd->code=CD_DIVCONST;
  cd->arg=k;
}
func zadd()
{
  var *scode:cd;
  cd=cg_getitem(ccg);
  cd->code=CD_ADD2REGS;
}
func zsub()
{
  var *scode:cd;
  cd=cg_getitem(ccg);
  cd->code=CD_SUB2REGS;
}
func neg()
{
  var *scode:cd;
  cd=cg_getitem(ccg);
  cd->code=CD_NEG;
}
func zeq()
{
  var *scode:cd;
  cd=cg_getitem(ccg);
  cd->code=CD_EQ;
}
func zne()
{
  var *scode:cd;
  cd=cg_getitem(ccg);
  cd->code=CD_NEQ;
}
func zge()
{
  var *scode:cd;
  cd=cg_getitem(ccg);
  cd->code=CD_ZGE;
}
func uge()
{
  var *scode:cd;
  cd=cg_getitem(ccg);
  cd->code=CD_UGE;
```

```
}
func ule()
{
  var *scode:cd;
  cd=cg_getitem(ccg);
  cd->code=CD_ULE;
}
func zle()
{
  var *scode:cd;
  cd=cg_getitem(ccg);
  cd->code=CD_ZLE;
}
func ult()
{
  var *scode:cd;
  cd=cg_getitem(ccg);
  cd->code=CD_ULT;
}
func zlt()
{
  var *scode:cd;
  cd=cg_getitem(ccg);
  cd->code=CD_ZLT;
}
func ugt()
{
  var *scode:cd;
  cd=cg_getitem(ccg);
  cd->code=CD_UGT;
}
func zgt()
{
  var *scode:cd;
  cd=cg_getitem(ccg);
  cd->code=CD_ZGT;
}
func zor()
{
  var *scode:cd;
  cd=cg_getitem(ccg);
  cd->code=CD_BOR2REGS;
}
func zxor()
{
  var *scode:cd;
  cd=cg_getitem(ccg);
  cd->code=CD_BXOR2REGS;
}
```

```
func zand()
{
  var *scode:cd;
  cd=cg_getitem(ccg);
  cd->code=CD_BAND2REGS;
}
func asr()
{
  var *scode:cd;
  cd=cg_getitem(ccg);
  cd->code=CD_ASR;
}
func asl()
{
  var *scode:cd;
  cd=cg_getitem(ccg);
  cd->code=CD_SHL;
}
func lnot()
{
  var *scode:cd;
  cd=cg_getitem(ccg);
  cd->code=CD_LNOT;
}
func bnot()
{
  var *scode:cd;
  cd=cg_getitem(ccg);
  cd->code=CD_BNOT;
}
func testjump(label:int)
{
  var *scode:cd;
  cd=cg_getitem(ccg);
  cd->code=CD_TESTJUMP;
  cd->arg=label;
}
func testnejump(label:int)
{
  var *scode:cd;
  cd=cg_getitem(ccg);
  cd->code=CD_TESTNEJUMP;
  cd->arg=label;
}
func jump(label:int)
{
  var *scode:cd;
  cd=cg_getitem(ccg);
  cd->code=CD_JUMP;
```

```
   cd->arg=label;
}
func zcall(sname:*char)
{
   var *scode:cd;
   cd=cg_getitem(ccg);
   cd->code=CD_ZCALL;
   cd->str=strdyn(sname);
}
func cmodstk(k:int)
{
   var *scode:cd;
   cd=cg_getitem(ccg);
   cd->code=CD_MODSTK;
   cd->arg=k;
}
func zenter()
{
   var *scode:cd;
   cd=cg_getitem(ccg);
   cd->code=CD_STKENTER;
}
func zleave()
{
   var *scode:cd;
   cd=cg_getitem(ccg);
   cd->code=CD_STKLEAVE;
}
func clab(int label)
{
   var *scode:cd;
   cd=cg_getitem(ccg);
   cd->code=CD_LAB;
   cd->arg=label;
}
func cloadlita(offs:int)
{
   var *scode:cd;
   cd=cg_getitem(ccg);
   cd->code=CD_LDLIT;
   cd->arg=offs;
}
func zldn(k:int)
{
   var *scode:cd;
   cd=cg_getitem(ccg);
   cd->code=CD_LDN;
   cd->arg=k;
}
```

```
func zlda(*char:name,int offset)
{
  var *scode:cd;
  cd=cg_getitem(ccg);
  cd->code=CD_LDA;
  cd->str=strdyn(name);
  cd->arg=offset;
}
func zlea(offset:int)
{
  var *scode:cd;
  cd=cg_getitem(ccg);
  cd->code=CD_LEA;
  cd->arg=offset;
}
func zstow(*char:name,offset:int)
{
  var *scode:cd;
  cd=cg_getitem(ccg);
  cd->code=CD_STOW;
  cd->str=strdyn(name);
  cd->arg=offset;
}
func zstob(*char:name,offset:int)
{
  var *scode:cd;
  cd=cg_getitem(ccg);
  cd->code=CD_STOB;
  cd->str=strdyn(name);
  cd->arg=offset;
}
func zstlw(int offset)
{
  var *scode:cd;
  cd=cg_getitem(ccg);
  cd->code=CD_STLW;
  cd->arg=offset;
}
func zstlb(int offset)
{
  var *scode:cd;
  cd=cg_getitem(ccg);
  cd->code=CD_STLB;
  cd->arg=offset;
}
func zlbrw(offset:int)
{
  var *scode:cd;
  cd=cg_getitem(ccg);
```

```
  cd->code=CD_LBRW;
  cd->arg=offset;
}
func zlbrb(offset:int)
{
  var *scode:cd;
  cd=cg_getitem(ccg);
  cd->code=CD_LBRB;
  cd->arg=offset;
}
func zlbra(offset:int)
{
  var *scode:cd;
  cd=cg_getitem(ccg);
  cd->code=CD_LBRA;
  cd->arg=offset;
}
func zldw(*char:name,int offset)
{
  var *scode:cd;
  cd=cg_getitem(ccg);
  cd->code=CD_LDW;
  cd->str=strdyn(name);
  cd->arg=offset;
}
func zldb(*char:name,int offset)
{
  var *scode:cd;
  cd=cg_getitem(ccg);
  cd->code=CD_LDB;
  cd->str=strdyn(name);
  cd->arg=offset;
}
func zldlw(int offset)
{
  var *scode:cd;
  cd=cg_getitem(ccg);
  cd->code=CD_LDLW;
  cd->arg=offset;
}
func zldlb(int offset)
{
  var *scode:cd;
  cd=cg_getitem(ccg);
  cd->code=CD_LDLB;
  cd->arg=offset;
}
func zstow2(int offset)
{
```

```
   var *scode:cd;
   cd=cg_getitem(ccg);
   cd->code=CD_STOW2;
   cd->arg=offset;
}
func zstob2(int offset)
{
   var *scode:cd;
   cd=cg_getitem(ccg);
   cd->code=CD_STOB2;
   cd->arg=offset;
}

func tstcg()
{
   var scodegen:cg;
   cg_init(&cg);
   var *scode:cd;
   cd=cg_getitem(&cg);
   cd->code=CD_LAB;
   cd->arg=19;
   cd=cg_getitem(&cg);
   cd->code=CD_ZCALL;
   cd->str=strdyn("function2");
   cd=cg_getitem(&cg);
   cd->code=CD_TESTJUMP;
   cd->arg=19;
   cd=cg_getitem(&cg);
   cd->code=CD_TESTNEJUMP;
   cd->arg=20;
   cg_print(&cg);
   cg_done(&cg);
}
var **char:regnames;
var *scodegen:ccg;
var scodegen:ccglb;
func icodegen()
{
   /*fprintf(stderr,"icodegen()\n");*/
   chkmem(regnames=calloc(4,sizeof(*char)));
   regnames[RG_A]="%eax";
   regnames[RG_B]="%ebx";
   regnames[RG_C]="%ecx";
   regnames[RG_D]="%edx";
   ccg=&ccglb;
   cg_init(ccg);
   /*fprintf(stderr,"ccg=%d\n",ccg);*/
}
func dcodegen()
```

```
{
  /*fprintf(stderr,"dcodegen()\n");*/
  /*fprintf(stderr,"ccg=%d\n",ccg);*/
  /*fprintf(stderr,"ccg.codeptr=%d\n",ccg->codeptr);*/
  cg_print(ccg);
  cg_done(ccg);
  if(regnames)
  free(regnames);
}
```

## 2.3   grph.e

```
/*                    -*- C -*-                              */
/*            The language compiler by E.V., (C) 2003               */
/* graphing module */
/*
  This program is free software; you can redistribute it and/or modify
  it under the terms of the GNU General Public License as published by
  the Free Software Foundation; either version 2 of the License, or
  (at your option) any later version.

  This program is distributed in the hope that it will be useful,
  but WITHOUT ANY WARRANTY; without even the implied warranty of
  MERCHANTABILITY or FITNESS FOR A PARTICULAR PURPOSE.  See the
  GNU General Public License for more details.

  You should have received a copy of the GNU General Public License
  along with this program; if not, write to the Free Software
  Foundation, Inc., 59 Temple Place, Suite 330, Boston, MA  02111-1307  USA
*/
#include "tlangc.he"
var *int:grphxof;
struct ssq
{
  [NAMESIZE]char name;
  int x,y;
  *snamelist nmlst;
  func draw;
};
#define SQNUM 300
var grsq:[SQNUM]ssq;
var grsqbx,grsqby:int;
var int:grsqptr;
var grsqx,grsqy,grsqstepx,grsqstepy:int;
func grsqadd(*char:name,*snamelist:nmlst)
{
  if(grsqptr>=SQNUM){error("graphing table full");return;}
  var int:i;
  strcp(grsq[grsqptr].name,name);
  grsq[grsqptr].nmlst=nmlst;
```

```
  grsq[grsqptr].x=grsqx;
  grsq[grsqptr].y=grsqy;
  grsqx=grsqx+grsqstepx;
  grsqy=grsqy+grsqstepy;
  if(grsqstepx>0&&grsqx>/*-grsqy*/grsqbx)
  {grsqstepy=grsqstepx;grsqstepx=0;grsqbx=grsqbx/*+11*/;}
  else if(grsqstepx<0&&-grsqx>/*grsqy*/grsqbx)
  {grsqstepy=grsqstepx;grsqstepx=0;grsqbx=grsqbx+11;}
  else if(grsqstepy>0&&grsqy>/*grsqx*/grsqby)
  {grsqstepx=-grsqstepy;grsqstepy=0;grsqby=grsqby/*+11*/;}
  else if(grsqstepy<0&&-grsqy>grsqby/*-grsqx*/)
  {grsqstepx=-grsqstepy;grsqstepy=0;grsqby=grsqby+11;}

  return grsqptr++;
}
func grsqfind(*char s)
{
  var int:i;
  for(i=0;i<grsqptr;i++)
  if(strid(grsq[i].name,s))
     return i;
  return i;
}
method ssq.draw()
{
  fprintf(grphxof,"set arrow from %d,%d to %d,%d nohead\n",x-5,y-5,x+5,y-5);
  fprintf(grphxof,"set arrow from %d,%d to %d,%d nohead\n",x+5,y-5,x+5,y+5);
  fprintf(grphxof,"set arrow from %d,%d to %d,%d nohead\n",x+5,y+5,x-5,y+5);
  fprintf(grphxof,"set arrow from %d,%d to %d,%d nohead\n",x-5,y+5,x-5,y-5);
  fprintf(grphxof,"set label \"%s\" at %d,%d center font\"Helvetica,5\"\n",name,x,y);
}
func printgraph()
{
  var of:*int;
  if(!(of=fopen(graphname,"w")))
  {
    fprintf(stderr,"can't open graph file %s\n",graphname);
    return;
  }
  grsqbx=0;
  grsqby=0;
  grphxof=of;
  grsqptr=0;
  grsqx=grsqy=0;
  grsqstepx=11;grsqstepy=0;
  fprintf(grphxof,"set terminal postscript color solid\n");
  fprintf(grphxof,"set noborder\n");
  fprintf(grphxof,"set noxtics\n");
  fprintf(grphxof,"set noytics\n");
```

```
fprintf(grphxof,"set nokey\n");
var *ssymlist:lst;
for(lst=glbsymtab.lst;lst;lst=lst->next)
{
  if(lst->sym.nmlst)
  {
    var int :i;
    i=grsqadd(lst->sym.name,lst->sym.nmlst);
    /*grsq[i].draw();*/
  }
  if(0)fprintf(of,"%-10s\t%d\t%d\n {",lst->sym.name,
      lst->sym.line,
      lst->sym.sort);
  if(0&&lst->sym.nmlst)
  {
    var *snamenode:q;
    q=lst->sym.nmlst->lst;
    while(q)
    {
      fprintf(of,"%s",q->name);
      if(q->next)fprintf(of,",");
      q=q->next;
    }
  }
  if(0)fprintf(of,"}\n");
}
var int:i;
for(i=0;i<grsqptr;i++)
{
  var *snamenode:q;
  var int:j;
  for(q=grsq[i].nmlst->lst;q;q=q->next)
  {
    j=grsqfind(q->name);
    if(j<grsqptr)
    {
      fprintf(grphxof,"set arrow from %d,%d to %d,%d nohead lt 2 lw 0.2\n",
          grsq[i].x,grsq[i].y,grsq[j].x,grsq[j].y);
    }
  }
}
for(i=0;i<grsqptr;i++)
{
  grsq[i].draw();
}
fprintf(grphxof,"plot [%d:%d][%d:%d] \"dat0.dat\"",-grsqbx-11,grsqbx+11,
    -grsqby-31,grsqby+31);
fclose(of);
}
```

## 2.4   tlangc.he

```
/*                    -*- C -*-                              */
/*        The language compiler definitions by E.V., (C) 2003        */
/*
  This program is free software; you can redistribute it and/or modify
  it under the terms of the GNU General Public License as published by
  the Free Software Foundation; either version 2 of the License, or
  (at your option) any later version.

  This program is distributed in the hope that it will be useful,
  but WITHOUT ANY WARRANTY; without even the implied warranty of
  MERCHANTABILITY or FITNESS FOR A PARTICULAR PURPOSE.  See the
  GNU General Public License for more details.

  You should have received a copy of the GNU General Public License
  along with this program; if not, write to the Free Software
  Foundation, Inc., 59 Temple Place, Suite 330, Boston, MA  02111-1307  USA
*/
#define WORDSIZE 4
#define BYTESIZE 1
#define linesize 180
#define linemax linesize-1
#define mpmax linemax
#define stif 1
#define stwhile 2
#define streturn 3
#define stbreak 4
#define stcont 5
#define stasm 6
#define stexp 7
#define stfor 9
#define stdo 10
#define NAMESIZE 16
#define NAMEMAX 15

struct snamenode{
  [NAMESIZE]char name;
  *snamenode next;
  func init;
  func done;
};
struct snamelist{
  *snamenode lst;
  **snamenode front;
  func init;
  func done;
  func find;
```

```
  func addm;
  func dump;
};
struct ssym{
  [NAMESIZE]char name;
  int sort;
  int dfd;
  int offset;
  int type;
  int line;
  *snamelist nmlst;
  func init;
  func done;
};
struct ssymlist{
  ssym sym;
  *ssymlist next;
};
struct ssymtab{
  *ssymlist lst;
  **ssymlist front;
};
var extern glbsymtab,locsymtab:ssymtab;

#define S_FUNC 1
#define S_VARG 2
#define S_VARL 3
#define T_INT 1
#define T_CHAR 2
#define T_INTP 3
#define T_CHARP 5
#define F_TYPE 6 /*first user-defined type*/

#define NUMTYP 50
#define MAXTYP 49
#define TYPNMTBS 800
#define V_PTR 1
#define V_ARR 2
#define V_STR 3
#define V_FND 4
#define V_FNC 5
struct styp{
  [NAMESIZE]char name;
  int sort;
  int type;
  int dim;
  int size;
  int ptr;
};
```

```
#define NMSTRELE 50
/*#define STRNMTBS 800*/
struct sfield{
  [NAMESIZE]char name;
  int type;
  int offset;
  int next;
};

#define litstksz 30000
#define litstknu 15

#define STSIZE 8000
#define STMAX 7999

#define WQNUM 75
#define WQMAX 74

/* the expression processing */
#define OP_LEAF    0
#define OP_COMMA   1
#define OP_ASSIGN  2
#define OP_COND    3
#define OP_LOR     4
#define OP_LAND    5
#define OP_BOR     6
#define OP_BXOR    7
#define OP_BAND    8
#define OP_EQ      9
#define OP_NEQ     10
#define OP_LE      11
#define OP_GE      12
#define OP_LT      13
#define OP_GT      14
#define OP_SHL     15
#define OP_SHR     16
#define OP_PLUS    17
#define OP_MINUS   18
#define OP_MUL     19
#define OP_DIV     20
#define OP_REM     21
#define OP_1PP     22
#define OP_1MM     24
#define OP_UPLUS   25
#define OP_UMINUS  26
#define OP_STAR    27
#define OP_ADDR    28
```

```
#define OP_LNOT   29
#define OP_BNOT   30
#define OP_2PP    31
#define OP_2MM    32
#define OP_SQ     33
#define OP_FUNC   34
#define OP_DOT    35
#define OP_LIST   36

#define L_NUM 1
#define L_ID  2
#define L_STR 3

/* pointed by %eax */
#define L_POI 4
/* pointed by (%esp) */
#define L_SP 5
/* on %eax */
#define L_ONREG 5
struct eleaf{
  int vid;
  int val;/* number or index in the symbol table*/
  *ssym idx;
};
struct enode{
  *enode l;
  *enode r;
  *enode third;
  int op;
  eleaf leaf;
  [NAMESIZE]char name;
};

struct elval{
  int sort; /* object, number, string -> L_ID,L_NUM,L_STR*/
  *ssym idx;  /* index in the symbol table, if object */
  int offset; /*offset in a compound object */
  int typ;    /* type */
  int val;
  [NAMESIZE]char name;
};
/*return value: 0: on register, 1: not on register */

var extern *char:graphname;
var extern stderr,stdin,stdout:*int;
```

## 2.5   codegen.he

```
/*                 -*- C -*-                          */
/*          code generation header by E.V., (C) 2003          */
```

```
/*
  This program is free software; you can redistribute it and/or modify
  it under the terms of the GNU General Public License as published by
  the Free Software Foundation; either version 2 of the License, or
  (at your option) any later version.

  This program is distributed in the hope that it will be useful,
  but WITHOUT ANY WARRANTY; without even the implied warranty of
  MERCHANTABILITY or FITNESS FOR A PARTICULAR PURPOSE.  See the
  GNU General Public License for more details.

  You should have received a copy of the GNU General Public License
  along with this program; if not, write to the Free Software
  Foundation, Inc., 59 Temple Place, Suite 330, Boston, MA  02111-1307  USA
*/
#define CD_MODSTK      10 /* add(or subtract) a number to(from) %esp */
#define CD_STKENTER    11 /* new stack frame */
#define CD_STKLEAVE    12 /* leave stack frame */
#define CD_INCREG      13 /* increase reg */
#define CD_DECREG      14 /* decrease reg */
#define CD_ADD2REGS    15 /* addition */
#define CD_SUB2REGS    16 /* subtraction */
#define CD_BAND2REGS   17 /* bitwise AND */
#define CD_BOR2REGS    18 /* bitwise OR  */
#define CD_BXOR2REGS   19 /* bitwise XOR */
#define CD_SHL         20 /* << */
#define CD_SHR         21 /* >> unsigned */
#define CD_ASR         22 /* >> signed   */
#define CD_EQ          23 /* == */
#define CD_NEQ         24 /* != */
#define CD_ZGE         25 /* >= signed */
#define CD_UGE         26 /* >= unsigned */
#define CD_ZLE         27 /* <= signed */
#define CD_ULE         28 /* <= unsigned */
#define CD_ZLT         29 /* < signed */
#define CD_ULT         30 /* < unsigned */
#define CD_ZGT         31 /* > signed */
#define CD_UGT         32 /* > unsigned */
#define CD_BNOT        33 /* ~ bitwise */
#define CD_LNOT        34 /* ! logical */
#define CD_NEG         35 /* - unary */
#define CD_TESTNEJUMP 36 /*  */
#define CD_ZCALL       37 /* call */
#define CD_LAB         38 /* label */
#define CD_MUL2REGS    39
#define CD_DIV2REGS    40
#define CD_MULREG      41 /* multiply register by const */
#define CD_DIVCONST     42 /* divide by const*/
#define CD_IGNORE      43
```

```
#define CD_MOD2REGS  44
#define CD_PUSH      45
#define CD_POP       46
#define CD_RET       47
#define CD_JUMP      48
#define CD_LDLIT     49
#define CD_LDN       50
#define CD_LDA       51
#define CD_LEA       52
#define CD_STOW      53
#define CD_STOB      54
#define CD_STLW      55
#define CD_STLB      56
#define CD_LBRW      57
#define CD_LBRB      58
#define CD_LBRA      59
#define CD_LDW       60
#define CD_LDB       61
#define CD_LDLW      62
#define CD_LDLB      63
#define CD_STOW2     64
#define CD_STOB2     65
#define CD_TESTJUMP  66 /*  */

struct scode
{
  int code;
  int arg;
  int reg;
  *char str;
};
#define NCODEITEMS 150
struct scodegen
{
  int ncodeitems;
  int codeptr;
  *scode codes;
};

func cg_init(this:*scodegen);
func cg_done(this:*scodegen);
func cg_getitem(this :*scodegen);
func cd_init(code:*scode);
func cg_print(this:*scodegen);

#define RG_A 0
#define RG_B 1
#define RG_C 2
```

```
#define RG_D 3

func icodegen();
func dcodegen();
var **char extern: regnames;
var extern *scodegen:ccg;
var extern scodegen:ccglb;

var extern int :Zsp;
```

## 2.6   autodyn.e

```
/*                    -*- C -*-                              */
/*        Automatic allocation/deallocation by E.V., (C) 2003            */
/*
  This program is free software; you can redistribute it and/or modify
  it under the terms of the GNU General Public License as published by
  the Free Software Foundation; either version 2 of the License, or
  (at your option) any later version.

  This program is distributed in the hope that it will be useful,
  but WITHOUT ANY WARRANTY; without even the implied warranty of
  MERCHANTABILITY or FITNESS FOR A PARTICULAR PURPOSE.  See the
  GNU General Public License for more details.

  You should have received a copy of the GNU General Public License
  along with this program; if not, write to the Free Software
  Foundation, Inc., 59 Temple Place, Suite 330, Boston, MA  02111-1307  USA
*/
#define _DYNSIZE 100
var int:_dynsize;
var **char: _dyn_ptr;
var int:_dynw;
func chkmem();
func initdyn()
{
  _dynsize=_DYNSIZE;
  chkmem(_dyn_ptr=calloc(_dynsize,sizeof(*char)));
  _dynw=0;
}
func dyncalloc(int:nmemb,int:size)
{
  if(_dynw>=_dynsize)
  {
    _dynsize=_dynsize+_DYNSIZE;
    chkmem(_dyn_ptr=realloc(_dyn_ptr,_dynsize*sizeof(*char)));
  }
  return _dyn_ptr[_dynw++]=chkmem(calloc(nmemb,size));
}
func autodynstr(*char:s)
```

```
{
  var int :l;
  var *char:res;
  l=strlen(s);
  res=dyncalloc(l+1,sizeof(char));
  strcp(res,s);
  return res;
}
var extern *int: stderr;
func donedyn()
{
  if(_dyn_ptr)
  {
    var int:i;
    i=_dynw;
    while(i)
    {
      free(_dyn_ptr[--i]);
      /*fprintf(stderr,"freed [%d]=%d\n",i,_dyn_ptr[i]);*/
    }
    free(_dyn_ptr);
  }
}
```

## 2.7   lpp1.e

```
/*      -*- C -*-                                                */
/***********************************************************************
                   This is the preprocessor
***********************************************************************/
/*
  This program is free software; you can redistribute it and/or modify
  it under the terms of the GNU General Public License as published by
  the Free Software Foundation; either version 2 of the License, or
  (at your option) any later version.

  This program is distributed in the hope that it will be useful,
  but WITHOUT ANY WARRANTY; without even the implied warranty of
  MERCHANTABILITY or FITNESS FOR A PARTICULAR PURPOSE.  See the
  GNU General Public License for more details.

  You should have received a copy of the GNU General Public License
  along with this program; if not, write to the Free Software
  Foundation, Inc., 59 Temple Place, Suite 330, Boston, MA  02111-1307  USA
*/
#define MAXINC 10
var extern stderr,stdin,stdout:*int;
var ifil,ofil:*int;
var ifiles:[MAXINC]*int;
var iptr:int;
```

```
var line,rline:[160]char;
var lptr,rlptr:int;
var isinp,iseof:int;
var strpool:[3000]char;
var strptr:int;
struct cmac{*char n; *char sub;};
var mactab:[200]cmac;
var macptr:int;
func strcp(d:*char,s:*char)
{
  while(*d++=*s++);
}
func ch()
{
  return line[lptr];
}
func gch()
{
  return line[lptr++];
}
func putres(c:int)
{
  if(rlptr>=159)
  {error("output buffer overflow");return 0;}
  return rline[rlptr++]=c;
}
func putmac(c:int)
{
  /*fprintf(stderr,"putmac:c=%d\n",c);*/
  if(strptr>=2999)
  {error("string pool full");return 0;}
  return strpool[strptr++]=c;
}
func getnewstr()
{
  if(strptr>=2999)
  {error("getnewstr:string pool full");return 0;}
  return strptr;
}
func addmac()
{
  var [16]char:sname;
  var int:k,l;
  if(!symname(sname))
  {error("wrong macro name");return;}
  if(macptr>=199)
  {error("mactab full");return;}
  sb();
  k=getnewstr();
```

```
    mactab[macptr].n=strpool+k;
    l=0;
    while(putmac(sname[l++]))
    ;
    k=getnewstr();
    mactab[macptr].sub=strpool+k;
    while(putmac(line[lptr++]));
    /*fprintf(stderr,"macro name=%s, sub=%s\n",mactab[macptr].n,
        mactab[macptr].sub);*/
    macptr++;
}
func strid(str1:*char,str2:*char)
{
    var int: k;
    k=0;
    while(str2[k])
    {if(str1[k]!=str2[k])return 0;k++;}
    if(str1[k])return 0;
    return 1;
}
func findmac(s:*char)
{
    var int: i;
    for(i=1;i<macptr;i++)
    if(strid(s,mactab[i].n))
        return i;
    return 0;
}
func symname(sname:*char)
{
    var int:k,l;var char: c;
    sb();
    if(!alpha(line[lptr]))return 0;
    for(k=l=0;an(line[lptr]);)
    {
        if(k<15)
        {sname[l++]=line[lptr++];k++;}
        else
        {k++;lptr++;}
    }
    sname[l]=0;
    return 1;
}
func amatch(lit:*char,len:int)
{
    var int:k;
    sb();
    if(k=astreq(line+lptr,lit,len))
    {
```

```
    lptr=lptr+k;
    return 1;
  }
  return 0;
}
func error(p:*char)
{
  fprintf(stderr,"***** Error:%s\nline:%s\n",p,line);
}
func an(c:int)
{
  return((alpha(c))||(numeric(c)));
}
func numeric(c:int)
{
  return((c>='0')&&(c<='9'));
}
func alpha(c:int)
{
  return (((c>='a')&&(c<='z'))||((c>='A')&&(c<='Z'))||(c=='_'));
}
func insline()
{
  var int:k;
  if(!isinp)iseof=1;
  if(iseof)return;
  lptr=0;
  line[0]=0;
  while((k=fgetc(ifil))>0)
  {
    if((k==10)||(lptr>=158))
    {break;}
    line[lptr++]=k;
  }
  line[lptr]=0;
  if(k<0)
  {
    if(iptr>0)
    {
      fclose(ifil);
      ifil=ifiles[--iptr];
    }
    else
    isinp=0;
  }
  /*fprintf(stderr,"<<<<<<%s\n",line);*/
}
func isb(c:int)
{
```

```
  return (line[lptr]==32)||(line[lptr]==9)||
  (line[lptr]==10);
}
func sb()
{
  while((line[lptr]==32)||(line[lptr]==9)||
    (line[lptr]==10))
  lptr++;
}
func match(lit:*char)
{
  var int:k;
  sb();
  if(k=streq(line+lptr,lit))
  {lptr=lptr+k;return 1;}
  return 0;
}
func streq(str1:*char,str2:*char)
{
  var int:k;
  for(k=0;str2[k];k++)
  if(str1[k]!=str2[k])return 0;
  return k;
}
func astreq(str1:*char,str2:*char,len:int)
{
  var int: k;
  k=0;
  while(k<len)
  {
    if((str1[k])!=(str2[k]))break;
    if(!str1[k])break;
    if(!str2[k])break;
    k++;
  }
  if(an(str1[k]))return 0;
  if(an(str2[k]))return 0;
  return k;
}
func prep()
{
  rline[0]=0;
  rlptr=0;
  lptr=0;
  lptr=0;
  while(line[lptr])
  {
    var [16]char:sname;
    var int:k;
```

```
if(isb(line[lptr]))
  {putres(' ');lptr++;}
else if(symname(sname))
{
  if(k=findmac(sname))
  {
    var *char:p;
    p=mactab[k].sub;
    /*fprintf(stderr,"substituting:%s\n",p);*/
    while(*p)putres(*p++);
  }
  else
  {
    k=0;
    while(sname[k])putres(sname[k++]);
  }
}
else if(line[lptr]==39)
{
  putres(gch());
  while(ch()!=39)
  {
    if(!ch())
    {error("unterminated char const");break;}
    else if(ch()=='\\')
    {
      putres(gch());
      if(!ch())
      {
        if(iseof)
        break;
        insline();
        lptr=0;
      }
      else putres(gch());
    }
    else putres(gch());
  }
  if(ch()==39)
  putres(gch());
}
else if(line[lptr]=='"')
{
  /*     fprintf(stderr,"quoted string=%s\n",line+lptr);*/
  putres(gch());
  while(ch()!='"')
  {
    if(!ch())
    {error("unterminated string");break;}
```

```
        else if(ch()=='\\')
        {
          putres(gch());
          if(!ch())
          {
            if(iseof)
            break;
            insline();
            lptr=0;
          }
          else putres(gch());
        }
        else putres(gch());
      }
      if(ch()=='"')putres(gch());
    }
    else if((line[lptr]=='/')&&(line[lptr+1]=='*'))
    {
      lptr=lptr+2;
      while(!((ch()=='*')&&(line[lptr+1]=='/')))
      {
        if(!ch())
        {
          if(iseof)
          break;
          insline();
          lptr=0;
        }
        else
        gch();
      }
      if((ch()=='*')&&(line[lptr+1]=='/'))
      lptr=lptr+2;
    }
    else putres(gch());
}
putres(0);
strcp(line,rline);
lptr=0;
if(amatch("#define",7))
{
  addmac();
  rline[0]=0;
}
else if(lptr=0,amatch("#include",8))
{
  doinclude();
  rline[0]=0;
}
```

```
}
func doinclude()
{
  var [160]char:newn;
  var int:k,c;
  var int:delim;
  sb();
  if(ch()=='"'){delim='"';gch();}
  else if(ch()=='<'){delim='>';gch();}
  else delim=0;
  for(k=0;k<159&&ch()&&!isblank(ch())&&ch()!=delim;k++)
    {
      if(ch()==92)
        {
          gch();
          if(!ch())break;
          c=gch();
          if(c=='n')c=10;
          else if(c=='t')c=9;
          else if(c=='b')c=8;
          else if(c=='f')c=12;
          newn[k]=c;
        }
      else
        {
          c=gch();
          newn[k]=c;
        }
    }
  newn[k]=0;
  /*fprintf(stderr,"new name:%s\n",newn);*/
  if(iptr>=MAXINC-2)
  error("too many nested #include");
  else
    {
      ++iptr;
      if(!(ifiles[iptr]=ifil=fopen(newn,"r")))
        {
          error("error opening %s\n",newn);
          ifil=ifiles[--iptr];
        }
    }
}
func process()
{
  while(!iseof)
    {
      insline();
      prep();
```

```
      fputs(rline,ofil);fputc(10,ofil);
    }
}
func main(argc:int,argv:**char)
{
  var int:i,is_out,is_in;
  var *char: outn,inn;
  strptr=0;
  macptr=1;
  iseof=0;
  isinp=1;
  is_in=is_out=0;
  outn=0;
  inn=0;
  iptr=0;
  for(i=1;i<argc;i++)
  {
    if(!is_in)
    {inn=argv[i];is_in=1;}
    else if(!is_out)
    {outn=argv[i];is_out=1;}
  }
  if(inn)
  {
    if(!(ifil=fopen(inn,"r")))
    {fprintf(stderr,"err in");exit(1);}
  }
  else
  ifil=stdin;
  if(outn)
  {
    if(!(ofil=fopen(outn,"w")))
    {fprintf(stderr,"err out");exit(1);}
  }
  else
  ofil=stdout;
  ifiles[0]=ifil;
  var int:c;
  process();

  if(inn)fclose(ifil);
  if(outn)fclose(ofil);
  return 0;
}
```

## 2.8   langdrv.pl

```
#!/usr/bin/perl
use File::Basename;
```

```perl
$LPP='./lpp1';
$LANGC='./langc';
$outname='a.out';
$numArgs=$#ARGV+1;
@enames=();
@snames=();
print "number of arguments:$numArgs\n";
for($i=0;$i<$numArgs;$i++)
   {
   print "\$ARGV[$i]=$ARGV[$i]\n";
   if($ARGV[$i] eq '-o')
      {
      if($i>=$numArgs-1)
         {
         print "ending with \"-o\"\n";
         exit 1;
         }
      $outname=$ARGV[$i+1];
      $i+=1;
      }
   else
      {
      push(@enames,$ARGV[$i]);
      ($xname,$xpath,$xsuffix)=fileparse($ARGV[$i],'\..*');
      print "\$xname=$xname, \$xpath=$xpath, \$xsuffix=$xsuffix, \n";
      $tsname=$xpath . $xname . '.s';
      print "\$tsname=$tsname\n";
      push(@snames,$tsname);
      }
   }
print("\$0=$0\n");
print "\@enames=@enames\n";
print "\@snames=@snames\n";
print "\$outname=$outname\n";
#$tt= $outname . "asdf\n";
#print $tt;
$nfiles=$#enames+1;
print "$nfiles files to compile\n";
for($i=0;$i<$nfiles;$i++)
   {
   $thecmd="$LPP $enames[$i] | $LANGC > $snames[$i]";
   print $thecmd . "\n";
   `$thecmd`;
   }
$slist="@snames";
print $slist . "\n";
$newcmd="gcc -o $outname @snames ";
print $newcmd . "\n";
`$newcmd`;
```